\documentclass[iop,twocolumn,tighten,numberedappendix,twocolappendix,revtex4]{emulateapj}
\usepackage{graphicx}
\usepackage{amsmath}
\usepackage{amssymb}
\usepackage{lipsum}
\usepackage{xcolor}

\renewcommand\ion[2]{#1$\;${\scshape{#2}}}

\graphicspath{{./}{plots/}}

\shorttitle{AS2UDS FIRRC}
\shortauthors{Algera et al.}

\begin{document}

\title{An ALMA Survey of the SCUBA-2 Cosmology Legacy Survey UKIDSS/UDS Field: The Far-infrared/Radio correlation for High-redshift Dusty Star-forming Galaxies}

\author{H.\,S.\,B.\,Algera\altaffilmark{1}}
\altaffiltext{1}{Leiden Observatory, Leiden University, P.O. Box 9513, 2300 RA Leiden, the Netherlands}
\email{algera@strw.leidenuniv.nl}

\author{I.\,Smail\altaffilmark{2}}
\altaffiltext{2}{Centre for Extragalactic Astronomy, Durham University, Department of Physics, South Road, Durham, DH1 3LE, UK}

\author{U.\,Dudzevi{\v{c}}i{\={u}}t{\.{e}}\altaffilmark{2}}

\author{A.\,M.\,Swinbank\altaffilmark{2}}

\author{S.\,Stach\altaffilmark{2}}

\author{J.\,A.\,Hodge\altaffilmark{1}} 

\author{A.\,P.\,Thomson\altaffilmark{3}}
\altaffiltext{3}{The University of Manchester, Oxford Road, Manchester, M13 9PL, UK}

\author{O.\,Almaini\altaffilmark{4}}
\altaffiltext{4}{School of Physics and Astronomy, University of Nottingham, University Park, Nottingham, NG7 2RD, UK}

\author{V.\,Arumugam\altaffilmark{5}}
\altaffiltext{5}{Institut de Radioastronomie Millim\'{e}trique, 300 rue de la Piscine, Domaine Universitaire, 38406 Saint Martin d’H\`{e}res, France}

\author{A.\,W.\,Blain\altaffilmark{6}}
\altaffiltext{6}{Department of Physics and Astronomy, University of Leicester,University Road, Leicester LE1 7RH, UK}

\author{G.\,Calistro-Rivera\altaffilmark{7}}
\altaffiltext{7}{European Southern Observatory, Karl-Schwarzchild-Strasse 2, 85748, Garching bei M\"{u}nchen, Germany}

\author{S.\,C.\,Chapman\altaffilmark{8}}
\altaffiltext{8}{Department of Physics and Atmospheric Science, Dalhousie, Halifax, NS B3H 4R2, Canada}

\author{C.-C\,Chen\altaffilmark{9}}
\altaffiltext{9}{Academia Sinica Institute of Astronomy and Astrophysics, No. 1, Sec. 4, Roosevelt Rd., Taipei 10617, Taiwan}

\author{E.\,da Cunha\altaffilmark{10,11,12}}
\altaffiltext{10}{International Centre for Radio Astronomy Research, University of Western Australia, 35 Stirling Hwy, Crawley, WA 6009, Australia}
\altaffiltext{11}{Research School of Astronomy and Astrophysics, The Australian National University, Canberra, ACT 2611, Australia}
\altaffiltext{12}{ARC Centre of Excellence for All Sky Astrophysics in 3 Dimensions (ASTRO 3D)}

\author{D.\,Farrah\altaffilmark{13,14}}
\altaffiltext{13}{Department of Physics and Astronomy, University of Hawaii, 2505 Correa Road, Honolulu, HI 96822, USA}
\altaffiltext{14}{Institute for Astronomy, 2680 Woodlawn Drive, University of Hawaii, Honolulu, HI 96822, USA}

\author{S.\,Leslie\altaffilmark{1}}

\author{D.\,Scott\altaffilmark{15}}
\altaffiltext{15}{ Department of Physics and Astronomy, University of British Columbia, 6224 Agricultural Road, Vancouver, BC V6T 1Z1, Canada}

\author{D.\,van der Vlugt\altaffilmark{1}}

\author{J.\,L.\,Wardlow\altaffilmark{16}}
\altaffiltext{16}{Physics Department, Lancaster University, Lancaster, LA14YB, UK}

\author{P.\,van der Werf\altaffilmark{1}}

\begin{abstract}
We study the radio properties of 706 sub-millimeter galaxies (SMGs) selected at $870\,\mu$m with the Atacama Large Millimeter Array from the SCUBA-2 Cosmology Legacy Survey map of the Ultra Deep Survey field. We detect 273 SMGs at $>4\sigma$ in deep \emph{Karl G. Jansky} Very Large Array 1.4\,GHz observations, of which a subset of 45 SMGs are additionally detected in 610\,MHz Giant Metre-Wave Radio Telescope imaging. We quantify the far-infrared/radio correlation through parameter $q_\text{IR}$, defined as the logarithmic ratio of the far-infrared and radio luminosity, and include the radio-undetected SMGs through a stacking analysis. We determine a median $q_\text{IR} = 2.20\pm0.03$ for the full sample, independent of redshift, which places these $z\sim2.5$ dusty star-forming galaxies $0.44\pm0.04\,$dex below the local correlation for both normal star-forming galaxies and local ultra-luminous infrared galaxies (ULIRGs). Both the lack of redshift-evolution and the offset from the local correlation are likely the result of the different physical conditions in high-redshift starburst galaxies, compared to local star-forming sources. We explain the offset through a combination of strong magnetic fields ($B\gtrsim0.2\,$mG), high interstellar medium (ISM) densities and additional radio emission generated by secondary cosmic rays. While local ULIRGs are likely to have similar magnetic field strengths, we find that their compactness, in combination with a higher ISM density compared to SMGs, naturally explains why local and high-redshift dusty star-forming galaxies follow a different far-infrared/radio correlation. Overall, our findings paint SMGs as a homogeneous population of galaxies, as illustrated by their tight and non-evolving far-infrared/radio correlation.
\end{abstract}

\keywords{galaxies: evolution $--$ galaxies: high-redshift $--$ galaxies: starburst}



\section{Introduction}

The most vigorously star-forming galaxies in the Universe are known to be highly dust-enshrouded, and as such reprocess the bulk of the ultra-violet radiation associated with massive star formation to emission at rest-frame far-infrared (FIR) wavelengths. While in the local Universe these galaxies contribute little to cosmic star formation (e.g.,\ \citealt{blain2002}), early sub-millimeter surveys discovered they were orders of magnitude more numerous at high-redshift \citep{smail1997,hughes1998,barger1998}. Accordingly, these distant, dust-enshrouded galaxies were dubbed sub-millimeter galaxies (SMGs, \citealt{blain2002}). The sub-millimeter surveys leading to their discovery were limited in angular resolution, complicating the identification of counterparts to SMGs at other wavelengths. An effective way around this difficulty was provided by follow-up radio observations with high enough resolution allowing for a less ambiguous determination of the origin of the far-infrared emission (\citealt{ivison1998,smail2000,lindner2011,barger2012}). This approach relies on the close connection between the total infrared output and radio luminosity of star-forming galaxies that has been known to exist for decades \citep{vanderkruit1971,vanderkruit1973,dejong1985,helou1985,condon1992,yun2001,bell2003}. The existence of this far-infrared/radio correlation (FIRRC) is a natural outcome if galaxies are `calorimeters', as proposed initially by \citet{volk1989} and \citet{lisenfeld1996}. In this model, galaxies are fully internally opaque to the ultra-violet (UV) radiation arising from massive star formation, such that these UV-photons are reprocessed by dust in the galaxy's interstellar medium, and subsequently re-radiated in the far-infrared. For this reason, far-infrared emission is a robust tracer of recent ($\lesssim100\,$Myr, e.g., \citealt{kennicutt1998}) star formation, provided the galaxy is optically thick to UV-photons. Since these very same massive stars ($M_\star \sim 8 - 40\,$M$_\odot$, \citealt{heger2003}) end their lives in Type-II supernovae, the resulting energetic cosmic rays traverse through the galaxy's magnetic field and lose energy via synchrotron emission. Provided only a small fraction of cosmic rays escape the galaxy before cooling, a correlation between the far-infrared and radio emission of a star-forming galaxy naturally arises \citep{volk1989}.

The ubiquity and apparent tightness of this correlation across a wide range of galaxy luminosities allows for the use of radio emission as an indirect indicator of dust-obscured star formation, and as such it has been widely utilized to study the history of cosmic star formation (e.g., \citealt{haarsma2000,smolcic2009,karim2011,novak2017}). This application of the far-infrared/radio correlation at high redshift, however, requires a clear understanding of whether it evolves across cosmic time. From a theoretical point of view, such evolution is indeed expected. For example, the increased energy density of the cosmic microwave background (CMB) at high redshift is expected to suppress radio emission in star-forming galaxies, as cosmic rays will experience additional cooling from inverse Compton scattering off the CMB (e.g., \citealt{murphy2009b,lacki2010b}). The exact magnitude of this process, however, will depend on the magnetic field strengths of the individual galaxies, which -- especially at high redshift -- are poorly understood. From an observational perspective, significant effort has been undertaken to assess whether the far-infrared/radio correlation evolves throughout cosmic time. While a number of studies find no evidence for such evolution (e.g., \citealt{ivison2010b,sargent2010,mao2011,duncan2020}), some studies suggest redshift-evolution in the far-infrared/radio correlation in the opposite sense to what is expected theoretically \citep{ivison2010a,thomson2014,magnelli2015,delhaize2017,calistrorivera2017,ocran2020}, seemingly implying that high-redshift ($z\gtrsim1$) star-forming galaxies have increased radio emission (or, alternatively, decreased far-infrared emission) compared to their local counterparts.

The most obvious explanation of this apparent evolution is contamination of the observed radio luminosity by emission from an active galactic nucleus (AGN) in the galaxy (e.g., \citealt{murphy2009a}). While such emission is straightforward to identify for radio-loud AGN -- precisely because it drives a galaxy away from the FIRRC -- composite sources may exhibit only low-level AGN activity, making them difficult to distinguish from typical star-forming galaxies (e.g., \citealt{beswick2008,padovani2009,bonzini2013}). A major uncertainty of the applicability of the FIRRC is therefore one's ability to identify radio-AGN, which is generally more challenging at high redshift. An additional potential driver of apparent redshift-evolution of the FIRRC involves sample selection (e.g., \citealt{sargent2010}). Differences in the relative depths of the radio and far-infrared observations, if not properly taken into account, will result in a biased sample. Additionally, the sensitivity of radio- and FIR-surveys to galaxies at high redshift are typically substantially different. While (sub-)mm surveys are nearly uniformly sensitive to dust-obscured star-formation across a wide range of redshifts ($1 \lesssim z \lesssim 10$, \citealt{blain2002}) and predominantly select galaxies at $z\approx2-3$ (e.g., \citealt{chapman2005,dudzeviciute2019}) owing to the strong, negative $K$-correction, radio surveys instead suffer from a positive $K$-correction \citep{condon1992}, and therefore predominantly select sources around $z\sim1$ \citep{condon1989}. Evidently, such selection biases must be addressed in order to assess the evolution of the far-infrared/radio correlation in the early Universe.

The cleanest way of studying any evolution in the FIRRC is therefore to start from a sample where the selection is well understood, and where radio AGN are less of a complicating factor. For this purpose, we employ the ALMA\footnote{Atacama Large Millimeter/sub-millimeter Array} SCUBA-2 UDS survey (AS2UDS), which constitutes the largest, homogeneously selected, sample of SMGs currently available \citep{stach2019,dudzeviciute2019}. While the far-infrared/radio correlation has been studied using FIR-selected samples before (e.g., \citealt{ivison2010a,ivison2010b,thomson2014}), the extent to which it evolves with cosmic time has remained unclear, due to either the limited resolution of the far-infrared data, the modest available sample sizes, or biases in these samples. The more than 700 ALMA-detected SMGs from the AS2UDS survey improve upon these shortcomings, and hence allow for a detailed investigation of the far-infrared/radio correlation for strongly star-forming sources at high redshift. 

The structure of this paper is as follows. In Section \ref{sec:observations} we outline the sub-millimeter and radio observations of the AS2UDS sample. In Section \ref{sec:results}, we separate radio-AGN from our sample, and investigate the redshift evolution of the star-forming SMGs. In Section \ref{sec:discussion} we discuss our results in terms of the physical properties of SMGs. Finally, we present our conclusions in Section \ref{sec:conclusion}. Throughout this paper, we adopt a flat $\Lambda$-Cold Dark Matter cosmology, with $\Omega_\text{m} = 0.30$, $\Omega_\Lambda = 0.70$ and $H_0 = 70\,\text{km}\,\text{s}^{-1}\,\text{Mpc}^{-1}$. We further assume a \citet{chabrier2003} Initial Mass Function, quote magnitudes in the AB system, and define the radio spectral index $\alpha$ such that $S_\nu \propto \nu^\alpha$, where $S_\nu$ represents the flux density at frequency $\nu$. 

\section{Observations \& Methods}
\label{sec:observations}

\subsection{Submillimeter Observations}
The AS2UDS survey \citep{stach2019} constitutes a high-resolution follow-up with ALMA of SCUBA-2 $850\,\mu$m sources originally detected over the UKIDSS Ultra Deep Survey (UDS) field as part of the SCUBA-2 Cosmology Legacy Survey (S2CLS, \citealt{geach2017}). The parent single-dish sub-millimeter survey spans an area of $0.96\,\text{deg}^2$, to a median depth of $\sigma_{850} = 0.88\,$mJy beam$^{-1}$. All sources detected at a significance of $>4\sigma$ ($S_{850} \geq 3.6\,$mJy) were targeted with ALMA observations in Band 7 ($344\,$GHz or $870\,\mu$m) across four different Cycles (1, 3, 4, 5). As a result, the beam size of the data varies between $0\farcs15-0\farcs5$, though for source detection all images were homogenized to $0\farcs5$ FWHM. Further details of the survey strategy and data reduction are presented in \citet{stach2019}. The final sub-millimeter catalog contains 708 SMGs detected at $\geq4.3\sigma$ ($S_{870} > 0.9\,$mJy), with an estimated false-positive rate of 2\%. 

\subsection{Radio Observations}
The UDS field has been observed at 1.4\,GHz by the \emph{Karl G. Jansky} Very Large Array (VLA). These observations will be fully described in Arumugam et al.\ (in prep.) and are additionally briefly summarized in \citet{thomson2019} and \citet{dudzeviciute2019}. In short, the 1.4-GHz image consists of a 14-pointing mosaic, for a total integration time of $\sim160\,$hr, across multiple VLA configurations. The bulk of the data ($\sim110\,$hr) were taken in A-configuration, augmented by $\sim50\,$hr of observations in VLA B-array and $\sim1.5\,$hr in the DnC configuration. The final root-mean-square (RMS) noise in the map is nearly uniform, reaching $7\,\mu\text{Jy}\,\text{beam}^{-1}$ in the image centre, up to $10\,\mu\text{Jy}\,\text{beam}^{-1}$ near the mosaic edges. The resulting synthesized beam is well-described by an elliptical Gaussian with major and minor axes of, respectively, $1\farcs8$ and $1\farcs6$. The final flux densities have been corrected for bandwidth-smearing, to be described in detail in Arumugam et al.\ (in prep.), and are provided for the AS2UDS sources by \citet{dudzeviciute2019}. Overall, 706/708 SMGs fall within the 1.4-GHz radio footprint covering the UDS field. These sources form the focus of this work.

The UDS field has further been targeted at 610 MHz by the Giant Metre-Wave Telescope (GMRT) during 2006 February 3-6 and December 5-10. Details of the data reduction and imaging are provided in \citet{ibar2009}. In summary, the GMRT image comprises a three-pointing mosaic, with each pointing accounting for 12\,hr of observing time. The final RMS noise of the 610-MHz mosaic is $45\,\mu\text{Jy}\,\text{beam}^{-1}$ in the image centre, and reaches up to $80\,\mu\text{Jy}\,\text{beam}^{-1}$ near the edges, for a typical value of $65\,\mu\text{Jy}\,\text{beam}^{-1}$. The synthesized beam of the image is well described by a slightly elliptical Gaussian of size $6\farcs1 \times 5\farcs1$. A total of 689 SMGs fall within the footprint of the 610 MHz observations. Source detection was performed using {\sc{PyBDSF}} \citep{mohanrafferty2015}, down to a peak threshold of $4.0\sigma$, leading to the identification of a total of 853 radio sources, though only a small fraction of those are associated with AS2UDS sub-millimeter galaxies (Section \ref{sec:results}). Due to the large beam size, the counterparts of AS2UDS SMGs are unresolved at 610 MHz, and as such we adopt peak flux densities for all of them. We further verified that source blending is not an issue, as only 2\% of AS2UDS SMGs have more than one radio-detected source at 1.4\,GHz in their vicinity within a GMRT beam full-width half maximum. In addition, for a source to be detected at 610\,MHz, but not at 1.4\,GHz, requires an unphysically steep spectral index of $\alpha\approx-2.7$, very different from the typical radio spectral index of $\alpha\sim-0.80$ (\citealt{condon1992,ibar2010}, see also Section \ref{sec:results}). As a result, the VLA map is sufficiently deep that further confusion or flux boosting at 610\,MHz can also be ruled out when no radio counterpart is detected at 1.4\,GHz.

\subsection{Additional Multi-wavelength Data}
In order to investigate the physical properties of our SMG sample, it is crucial to obtain panchromatic coverage of their spectral energy distributions (SEDs). At UV, optical and near-infrared wavelengths, these SEDs are dominated by (dust-attenuated) stellar emission, which includes spectral features that are critical for obtaining accurate photometric redshifts. As SMGs are typically high-redshift in nature ($z\approx2-3$, e.g., \citealt{chapman2005,danielson2017}), these rest-frame wavelengths can be probed with near- and mid-infrared observations. The multi-wavelength coverage of the UDS field, as well as the association of counterparts to the SMG sample, is described in detail in \citet{dudzeviciute2019}, and further summarized in their Table 1, although we briefly repeat the key points here.

\citet{dudzeviciute2019} collated optical/near-infrared photometry for the AS2UDS SMGs from the 11$^\text{th}$ UDS data release (UKIDSS DR11, Almaini et al.\ in prep.). DR11 constitutes a $K$-band-selected photometric catalog covering an area of $0.8\,\text{deg}^2$. The $K$-band image reaches a $3\sigma$ depth of $25.7\,$mag, in $2''$ diameter apertures, and the resulting photometric catalog contains nearly 300,000 sources. This catalog further contains photometry in the $J$- and $H$-bands from the UKIRT WFCAM, as well as $Y$-band observations from VISTA/VIDEO, $BVRi'z'$-band photometry from Subaru/Suprimecam and $U$-band observations from the CFHT/Megacam survey. 

In total, 634 SMGs lie within the area covered by deep $K$-band imaging. The ALMA and $K$-band selected catalogs have been cross-matched using a radius of $0\farcs6$, resulting in 526/634 associations with an expected false-match rate of $3.5\%$. A significant number of SMGs, 17\%, are hence undetected even in deep $K$-band imaging (see Smail et al., in prep.). Further imaging in the infrared is provided by \emph{Spitzer}, in the four IRAC channels, as well as MIPS $24\,\mu$m, as part of the \emph{Spitzer} Legacy Program (SPUDS, PI: J.\ Dunlop). Upon adopting a conservative blending criterion where SMGs with nearby $K$-band detections are treated as upper limits (see \citealt{dudzeviciute2019} for details), 73\% of the SMGs covered by the IRAC maps are detected at $3.6\,\mu$m. In total, 48\% of SMGs are further detected at $24\,\mu$m. 

While the AS2UDS sample is, by construction, detected in the sub-millimeter at $870\,\mu$m, additional sampling of the long-wavelength dust continuum is crucial in order to obtain accurate far-infared luminosities, as well constraints on SMG dust properties, such as dust masses and temperatures. For this purpose, we employ observations taken with the PACS and SPIRE instruments aboard the \emph{Herschel} Space Observatory. To compensate for the coarse point spread function at these wavelengths and the resulting source blending, \citet{dudzeviciute2019} deblended the data following \citet{swinbank2014}, adopting ALMA, \emph{Spitzer}/MIPS $24\,\mu$m and 1.4 GHz observations as positional priors. Overall, 68\% of ALMA SMGs have a measured (potentially deblended) flux density in at least one of the PACS or SPIRE bands.

\subsection{SMG Redshifts and Physical Properties}
The redshift distributions, as well as numerous other physical properties of the AS2UDS SMGs, have been investigated by \citet{dudzeviciute2019}. For this, they employ the SED-fitting code {\sc{magphys}} \citep{dacunha2008,dacunha2015,battisti2019}, which is designed to fit the full UV-to-radio SED of star-forming galaxies. In order to self-consistently constrain the spectral energy distribution, {\sc{magphys}} employs an energy balance procedure, whereby emission in the UV, optical, and near-infrared is physically coupled to the emission at longer wavelengths by accounting for absorption and scattering by dust within the galaxy. The star-formation histories of individual galaxies are modeled as a delayed exponential function, following \citet{lee2010}, which corresponds to an initial linearly increasing star-formation rate, followed by an exponential decline. In addition, it allows for bursts to be superimposed on top this continuous star-formation history, during which stars are formed at a constant rate for up to $300\,$Myr. We note, however, that constraining the star-formation history and assigning ages by fitting to the broadband photometry of strongly dust-obscured galaxies is notoriously challenging (e.g, \citealt{hainline2011,michalowksi2012,simpson2014}). Further details of the {\sc{magphys}} analysis, including an extensive description of calibration and testing, are provided in \citet{dudzeviciute2019}. 

The latest extension of {\sc{magphys}}, presented in \citet{battisti2019}, further incorporates fitting for the photometric redshifts of galaxies. Accurate redshift information is crucial for a complete characterization of the SMG population, as any uncertainties on a galaxy's redshift will propagate into the error on derived physical quantities. Incorporating far-infrared data in the fitting can further alleviate degeneracies between optical colours and redshift, potentially allowing for a more robust determination of photometric redshifts \citep{battisti2019}. This is especially relevant for sub-millimeter galaxies, as these typically constitute an optically faint population.

In total, 44 AS2UDS SMGs have a measured spectroscopic redshift. \citet{dudzeviciute2019} compared the photometric redshifts (derived for both this SMG sub-sample, as well as for around $7000$ field galaxies in the UDS field with spectroscopic redshifts) to the existing spectroscopic ones, and find a photometric accuracy of $\Delta z / (1+z_\text{spec}) = -0.005 \pm 0.003$. Hence, the photometric redshifts provided by {\sc{magphys}} are in excellent agreement with the spectroscopic values. The typical uncertainty on the photometric redshift for the AS2UDS SMGs is $\Delta z \approx 0.25$.

Finally, various physical quantities are determined for the SMG sample via {\sc{magphys}}, including star-formation rates, mass-weighted ages, stellar and dust masses, as well as far-infrared luminosities. The accuracy of these values has been assessed by \citet{dudzeviciute2019} through comparing with simulated galaxies from {\sc{EAGLE}} \citep{schaye2015,crain2015,mcalpine2019}, where these properties are known a priori. The simulated and {\sc{magphys}}-derived values for the various physical parameters are typically in good agreement.\\

We caution that {\sc{magphys}} does not allow for any contribution from an AGN to the overall SED. In particular, emission from a mid-infrared power-law component, indicative of an AGN torus, may therefore result in slightly boosted FIR-emission. Such a mid-infrared power law is however not expected to contaminate the observed $870\,\mu$m (rest-frame $\sim250\,\mu$m) flux density (e.g., \citealt{lyu2017,xu2020}), and as such does not affect our sample selection. Therefore, we can quantify the typical contribution of the mid-infrared power-law to the total infrared luminosity for the AS2UDS SMGs. For this, we limit ourselves to the 442 SMGs at $z \leq 3.0$, following \citet{stach2019}, since above this redshift the criteria are prone to misclassifying dusty star-forming galaxies. This constitutes a total of 82 sources (12\% of the full AS2UDS sample). We find the median $8-1000\,\mu$m luminosity to be $\log_{10} L_\mathrm{FIR} = 12.54_{-0.04}^{+0.07}\,L_\odot$ and $\log_{10} L_\mathrm{FIR} = 12.33_{-0.03}^{+0.02}\,L_\odot$ for sources with and without a mid-infrared power-law, respectively. We therefore conclude that the typical AGN contribution to the total infrared luminosity is at most $\lesssim0.2\,$dex.

An additional diagnostic of an AGN is luminous X-ray emission. However, only one-third of the AS2UDS SMGs lie within the footprint of the available \emph{Chandra} X-ray imaging as part of the X-UDS survey (\citealt{kocevski2018}; see also \citealt{stach2019}). In particular, out of the 23 SMGs associated with strong X-ray emitters, 18 are additionally identified as AGN through their mid-infrared power-law emission. Therefore, when discussing AGN in the SMG population, we focus on the 82 mid-infrared-selected sources, which make up the bulk of the AGN in the AS2UDS sample. 

Studies of radio-selected samples have shown that AGN activity at radio wavelengths is often disjoint from AGN-related emission at X-ray and mid-infrared wavelengths (e.g., \citealt{delvecchio2017,algera2020}). In particular, \citet{algera2020} show that radio sources with X-ray and/or mid-infrared power-law emission fall onto the same far-infrared/radio correlation as ``clean'' star-forming sources. For this reason, we have decided to retain sources with non-radio AGN signatures in our sample (Section \ref{sec:results_agn}). In all relevant figures in this work, we however distinguish between ``clean SMGs'' and sources with a mid-infrared power-law signature via different plotting symbols. We additionally emphasize that our results are unaffected if these AGN are removed from the analysis entirely.

\subsection{Radio Stacking}
A comprehensive analysis of the far-infrared/radio correlation requires addressing any biases in the sample selection. In particular, the majority of AS2UDS sources are not detected in the $1.4$-GHz VLA map (about 60\%; Section \ref{sec:radio_properties}), yet their -- a priori unknown -- radio properties must still be included in the analysis. In this work, we employ a stacking technique in order to obtain a census of the typical radio properties of the AS2UDS sample. 

For the stacking, we create cutouts of $51\times51$ pixels ($18''\times18''$) from the 1.4 GHz radio map, centered on the precise ALMA positions of the AS2UDS sources. We average these cutouts together by taking the median value across each pixel. In order to properly account for the full SMG population, we stack both the radio-detected and -undetected SMGs together. Additionally, we stack empty regions within the image, away from radio sources, to create an ``empty'' stack indicative of the background and typical RMS-value (following e.g., \citealt{decarli2014}). We have verified that the RMS is reduced following a typical $1/\sqrt{N}$-behaviour, where $N$ is the number of sources being stacked. This indicates we are not significantly affected by confusion noise. We pass both the real and empty stacks to {\sc{PyBDSF}} \citep{mohanrafferty2015} to obtain peak, integrated, and aperture flux densities. We have run extensive simulations, using mock sources inserted into the image plane, to ascertain which flux density is the correct one to use. We elaborate on these simulations in Appendix \ref{app:stacking}, and will describe them in further detail in a future work (Algera et al.\ in prep.). The simulations show that integrated fluxes provide the most robust flux measurement for our data at moderate signal-to-noise ($\text{SNR}\gtrsim10$). In this work, we therefore use integrated fluxes obtained from {\sc{PyBDSF}}. The only exceptions are the GMRT 610-MHz stacks described in Section \ref{sec:results_qtir}, since due to the large beam size (about $5''$) all stacks are unresolved, and peak and integrated flux densities are consistent. For the GMRT stacks, we therefore adopt peak flux densities. 

In order to determine realistic uncertainties on the stacked flux densities, we perform a bootstrap analysis, whereby we repeat the procedure described above 100 times. This involves sampling SMGs from each bin with replacement, such that duplicate cutouts are allowed. In this way, the uncertainties on the final flux density reflect both the uncertainties on the photometry, as well as the intrinsic variation in the radio flux densities among the AS2UDS SMGs.

\section{Results}
\label{sec:results}

\subsection{Radio Properties of AS2UDS}
\label{sec:radio_properties}

\begin{figure}[!t]
    \centering
    \includegraphics[width=0.5\textwidth]{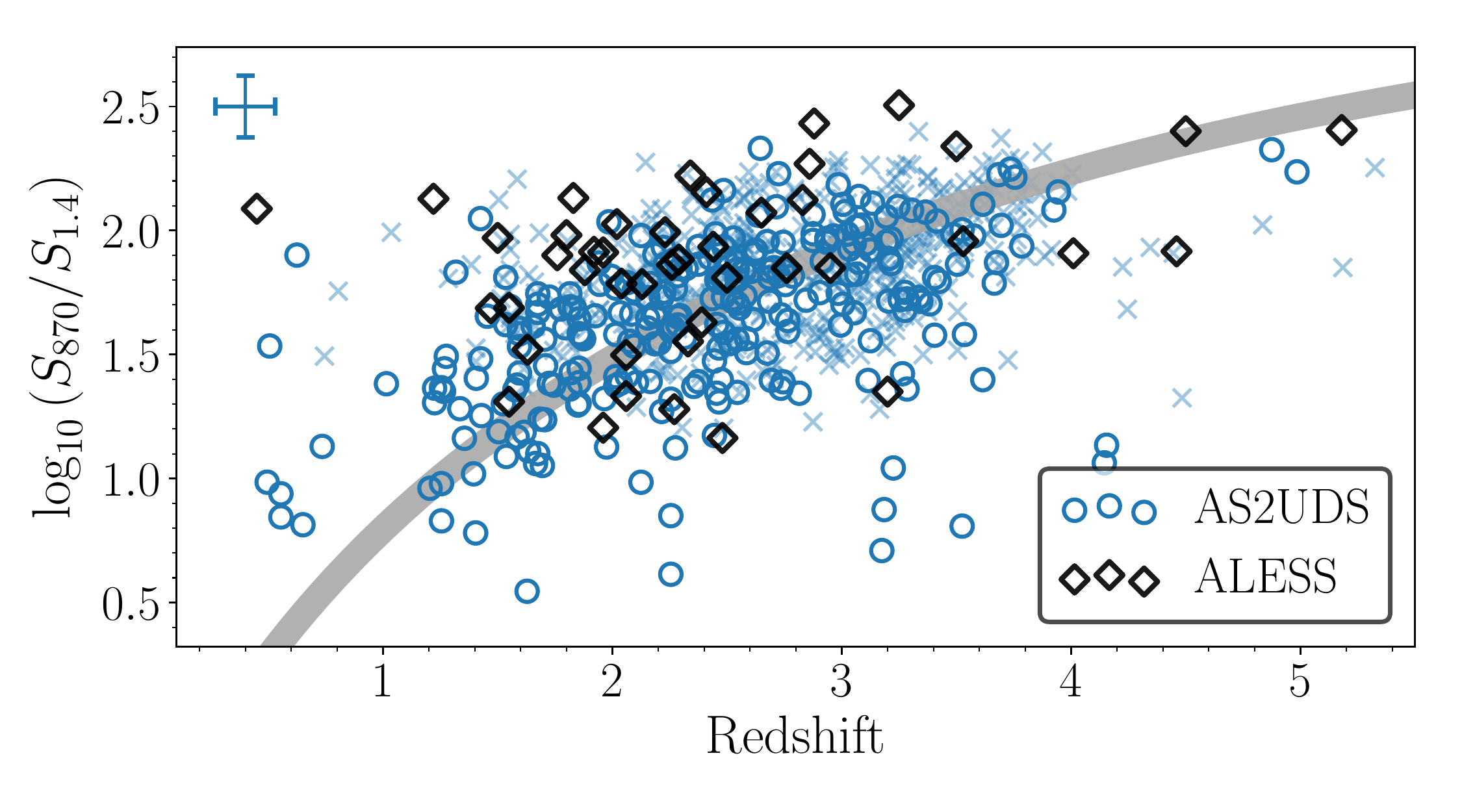}
    \caption{Ratio of the sub-millimeter to radio flux density as a function of redshift, for both the AS2UDS and ALESS samples. This ratio provides a crude proxy for redshift \citep{carilli1999}, as a result of the different typical $K$-corrections at $870\,\mu$m and 1.4 GHz. The expected trend with redshift is overlaid in gray, assuming a fixed far-infrared luminosity, dust emissivity and temperature, and FIRRC-parameter $q_\text{IR}$ (see text for details). In total, 273 (433 undetected) AS2UDS SMGs are detected at $\geq4\sigma$ (lower limits are shown as crosses) at 1.4 GHz, compared to 44 (32 undetected; not shown) for ALESS. The increase in sample size in comparison to ALESS constitutes nearly a factor of ten.}
    \label{fig:data}
\end{figure}

In total, 273 out of the 706 SMGs in the 1.4-GHz coverage of AS2UDS (39\%) can be cross-matched to a radio counterpart detected at $\geq4\sigma$ at 1.4 GHz, within a matching radius of $1\farcs6$ (chosen such that the fraction of false positives is $1\%$; \citealt{dudzeviciute2019}). This detection fraction is typical for high-redshift SMGs (e.g., \citealt{biggs2011,hodge2013}). We additionally detect 45 SMGs down to a $4\sigma$ threshold in the shallower 610 MHz observations. All of the sources detected in the 610 MHz map have a counterpart at 1.4 GHz, based on a cross-matching radius of $2\farcs0$. This is slightly larger than matching radius adopted for the VLA radio data to account for the coarser GMRT 610 MHz resolution, but still ensures a small false positive fraction of $\lesssim 0.1\%$.

We present the far-infrared and radio properties of the AS2UDS sample in Figure \ref{fig:data}, which shows the ratio of sub-millimeter to radio flux density for the AS2UDS SMGs as a function of redshift. As result of the different $K$-corrections in the far-infrared and radio, this ratio provides a crude proxy for redshift (e.g., \citealt{carilli1999}). The AS2UDS detections are consistent with the expected trend, plotted for a galaxy with a far-infrared luminosity of $10^{12.5} L_\odot$, which is typical for AS2UDS \citep{dudzeviciute2019}. This further assumes a fixed dust emissivity and temperature of $\beta = 1.8$ and $T_\text{dust} = 35\,$K, respectively, as well as a fixed radio spectral index of $\alpha = -0.8$ and a redshift-independent FIRRC, equal to the median value for AS2UDS (Section \ref{sec:results_qtir}). There is, however, substantial scatter around this trend, as may be expected from intrinsic variations in the dust and radio properties of our SMG sample.

Figure \ref{fig:data} further emphasizes the substantial increase in sample size that AS2UDS provides compared to the ALESS survey \citep{hodge2013,karim2013}. The latter constitutes an ALMA follow-up of SMGs originally identified in the Extended \emph{Chandra} Deep Field South as part of the LESS survey using the LABOCA bolometer \citep{weiss2009}. ALESS is similar to AS2UDS in terms of sample selection, and therefore provides the best means of comparison for this work. Additionally, the depth of both its far-infrared and radio observations closely match that of AS2UDS. In total, the ALESS survey covers 76 SMGs within its radio footprint \citep{thomson2014}. AS2UDS, therefore, constitutes a sample nearly ten times larger than ALESS. We compare the combined far-infrared and radio properties of the AS2UDS and ALESS samples in Section \ref{sec:comparison}. \\

\begin{figure*}[!t]
    \centering
    \hspace*{-1cm}\includegraphics[width=1.1\textwidth]{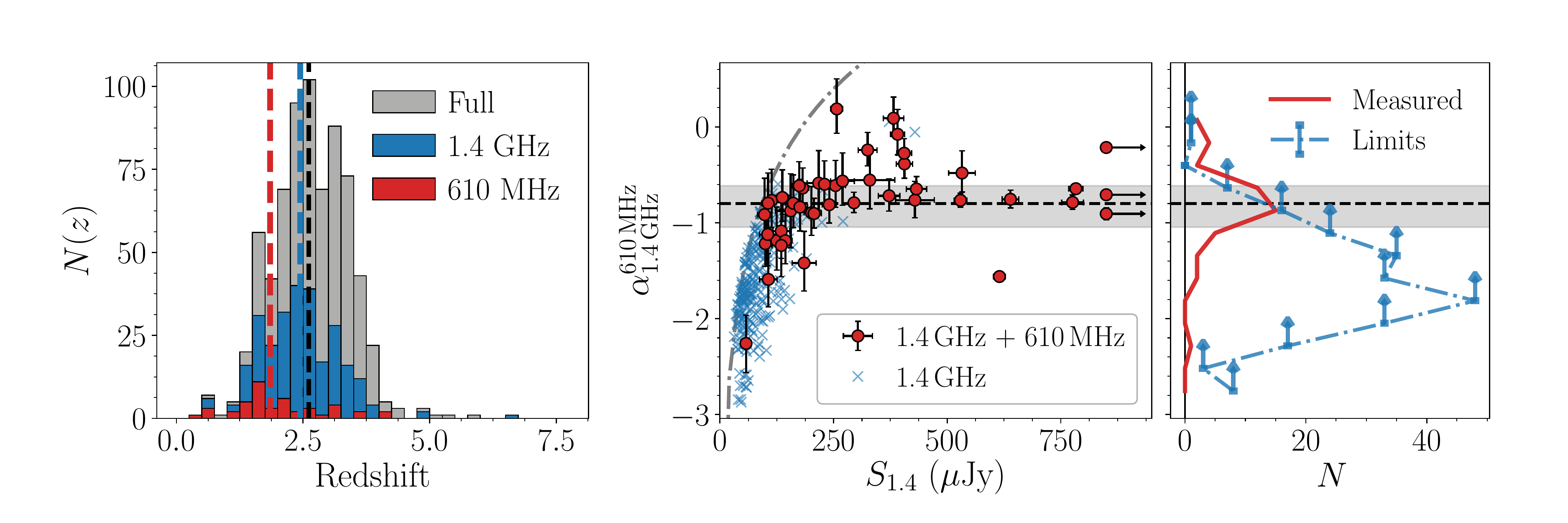}
    \caption{\textbf{Left:} Distribution of the radio-detected AS2UDS population as a function of redshift. The full AS2UDS sample is shown, as are the subset detected at $1.4\,$GHz, and those sources detected at both $610\,$MHz and $1.4\,$GHz. The vertical, dashed lines show the median redshift of these three populations. The radio-detected subset lies at a slightly lower redshift than the full AS2UDS sample, as a result of the different $K$-corrections for the typical FIR- and radio-detected populations. \textbf{Middle:} Radio spectral index as a function of the 1.4-GHz flux density. Sources with a $610\,$MHz detection, and hence with a measured spectral index, are highlighted. As expected, a large fraction of the lower limits on the spectral index corresponds to faint radio sources. The dash-dotted line indicates the shallowest spectral index these sources can have in order to be detected at both 610 and $1400\,$MHz, assuming the central RMS of $45\,\mu$Jy$\,\text{beam}^{-1}$ at $610\,$MHz. For the limits, we adopt a fixed spectral index of $\alpha=-0.80$ (dashed horizontal line). This value lies well within the $1\sigma$ uncertainty on the stacked spectral index we find for AS2UDS subset detected at $1.4\,$GHz but not at $610\,$MHz (gray shaded region). Three sources with $S_{1.4} > 1\,\text{mJy}$ lie outside the plotting limits, and are shown as the arrows placed on the right. \textbf{Right:} Distribution of spectral indices for the radio-detected AS2UDS sample, including direct measurements and lower limits. In this work, we mostly rely on a single radio detection at 1.4\,GHz, and hence adopt a fixed spectral index for the majority of the radio-detected SMG sample.}
    \label{fig:distributions}
\end{figure*}

We show the redshift distribution of the AS2UDS sources with radio detections in Figure \ref{fig:distributions} (left panel). As expected, the radio sources lie at a slightly lower redshift than the overall AS2UDS population, owing to the different $K$-corrections for typical radio and submillimeter detected sources. The median redshift of the 1.4-GHz detected subsample is $\langle z \rangle = 2.44_{-0.15}^{+0.04}$, while that of the 45 GMRT-detected sources is $\langle z \rangle = 1.85_{-0.21}^{+0.24}$, compared to $\langle z \rangle = 2.62_{-0.04}^{+0.06}$ for the full sample of AS2UDS SMGs \citep{dudzeviciute2019}.

We find a typical spectral index between 610 and 1400 MHz of $ \alpha = -0.77_{-0.03}^{+0.05}$, consistent with the typical radio spectrum of star-forming galaxies of $\alpha \approx -0.80$ (e.g., \citealt{condon1992,ibar2010}). Nevertheless, there is substantial variation in the spectra among the 45 sources detected at the two radio frequencies, with the $16^\text{th}$-$84^\text{th}$ percentile range spanning $\alpha \in [-1.19, -0.48]$. This range is wider than the variation expected based on the typical uncertainty on the spectral index of $\sim0.24\,$dex, indicating that at least some of this scatter is intrinsic variation in the radio spectral indices. The full distribution of spectral indices, including lower limits for sources detected solely at 1.4 GHz, is shown in the middle and right panels of Figure \ref{fig:distributions}. These limits were calculated by adopting four times the local RMS noise at the position of the radio source as an upper limit on the GMRT flux density. It is evident that most of the resulting lower limits on the spectral index are not very constraining, due to the limited depth of the 610\,MHz data. In order to assign a spectral index to the entire radio-detected population, we median stack all AS2UDS SMGs detected solely at 1.4\,GHz in both radio maps (225 sources within both the VLA and GMRT footprints). The typical stacked $610-1400\,$MHz spectral index is then found to be $\alpha = -0.81_{-0.23}^{+0.20}$. This value is consistent with the median spectral index obtained for the AS2UDS subsample having two radio detections, as well as with the typically assumed value of $\alpha = -0.80$ for SMGs. For ease of comparison to the literature, we will therefore adopt a fixed $\alpha=-0.80$ for all AS2UDS SMGs detected only in the 1.4-GHz map. We note that, while the beam size of our GMRT observations is significantly larger than that of the VLA data, the typical low-frequency radio sizes of SMGs are $\sim0.5-1.5''$ \citep{miettinen2017,jimenez-andrade2019,thomson2019}, similar to the synthesized beam at 1.4\,GHz. As such, this is much smaller than the largest angular scale to which we are sensitive with our data of $\sim120''$ based on the $\sim50\,$h of data taken in the VLA B-array configuration.\footnote{The largest angular scale in A-array, accounting for two-thirds of the observation time, equals $36''$, still significantly ($\sim40\times$) larger than the typical radio sizes of SMGs.} \citet{thomson2019} have further empirically verified the robustness of the theoretical largest angular scale, and as such, we do not expect to miss any diffuse emission at 1.4\,GHz. Our spectral index measurements are therefore unaffected by the differing resolutions of our radio observations (see also \citealt{gim2019}). We further discuss the spectral indices of the AS2UDS sample in Section \ref{sec:results_qtir}.

Given these spectral indices for the radio-detected SMG subsample, we calculate the luminosity at a rest-frame frequency of $\nu = 1.4\,$GHz as 

\begin{align}
	L_\text{1.4} = \frac{4\pi D_L^2}{(1+z)^{1 + \alpha}} S_\text{1.4} \ .
	\label{eq:luminosity}
\end{align}

Here $D_L$ is the luminosity distance to a source at redshift $z$, and $S_\text{1.4}$ is its flux density at the observer-frame frequency of 1.4 GHz. Note that our 1.4-GHz radio observations probe a typical rest-frame frequency of $\nu \sim 5$ GHz, for a source at the median AS2UDS redshift. Adopting the $16^\text{th}$ or $84^\text{th}$ percentile of our spectral index distribution for the $K$-correction (instead of $\alpha = -0.80$) leads to a typical difference of a factor of $1.5\times$ in the rest-frame 1.4-GHz radio luminosity. For SMGs without a radio counterpart, we adopt $4\times$ the local RMS-noise in the 1.4\,GHz map and a fixed spectral index of $\alpha=-0.80$ in order to calculate the corresponding upper limit on the radio luminosity. The far-infrared luminosities for the AS2UDS sample, obtained via {\sc{magphys}}, are determined in the wavelength range $8-1000\,\mu$m, and allow us to define the parameter $q_\text{IR}$ characterizing the far-infrared/radio correlation. Following e.g., \citet{condon1991b,bell2003,magnelli2015,calistrorivera2017}, we define it as:

\begin{align}
	q_\text{IR} = \log_{10} \left( \frac{L_\text{FIR}}{3.75 \times 10^{12}\text{ W}} \right) - \log_{10} \left( \frac{L_\text{1.4}}{\text{W Hz}^{-1}} \right)  \ .
	\label{eq:qTIR}
\end{align}

Here, the FIR-luminosity $L_\text{FIR}$ is normalized such that $q_\text{IR}$ is dimensionless. For the full radio-detected subsample, we find a median $ q_\text{IR} = 2.10 \pm 0.02$. Had we neglected the 610-MHz data and instead assumed a fixed spectral index of $\alpha = -0.80$, we would obtain a similar value of $ q_\text{IR} = 2.11 \pm 0.02$. This value is lower than what is found for local, typically less strongly star-forming galaxies of $ q_\text{IR} = 2.64 \pm 0.02$ \citep{bell2003}. However, it is similar to the values found by \citet{kovacs2006} and \citet{magnelli2010} of respectively $q_\text{IR} = 2.07 \pm 0.09$ and $ q_\text{IR} = 2.17 \pm 0.19$ for $z\approx2$ radio-detected dusty star-forming galaxies, although other studies of SMGs find typical values for $q_\text{IR}$ that are more similar to the local correlation (e.g., \citealt{sargent2010,ivison2010a}). We emphasize, however, that the average value of $q_\text{IR}$ for any given sample is highly dependent on its selection, and the relative depths of the far-infrared and radio observations. Therefore, we compare with the results from the ALESS survey by \citet{thomson2014} in more detail in Section \ref{sec:comparison}, as both its sub-millimeter selection and radio coverage at 1.4 GHz are similar to that of AS2UDS. \\

In the following sections, we will study the far-infrared/radio correlation for two samples. First of all, we utilize all SMGs within the redshift range $1.5 \leq z \leq 4.0$, totalling 659 sources (93\% of the entire AS2UDS sample). We limit ourselves to this redshift range to provide a more uniform selection of SMGs \citep{dudzeviciute2019}, and will refer to this sample as the ``full AS2UDS sample''. Secondly, we follow \citet{dudzeviciute2019} and focus on the 133 SMGs at $1.5 \leq z \leq 4.0$ within the luminosity range $L_\text{FIR} = 4 - 7 \times 10^{12}\ L_\odot$ with at least one \emph{Herschel}/SPIRE detection. By restricting ourselves to this luminosity range, we ensure the sample is complete with respect to the SPIRE detection limits. As such, we retain a subsample complete in far-infrared luminosity, but with better constraints on its dust properties, due to the additional sampling of the far-infrared SEDs. Following \citet{dudzeviciute2019}, we will refer to this sample as the ``luminosity-limited sample''.

\subsection{AGN in AS2UDS}
\label{sec:results_agn}

\begin{figure*}[!t]
    \centering
    \hspace*{-0.45cm}\includegraphics[width=1.05\textwidth]{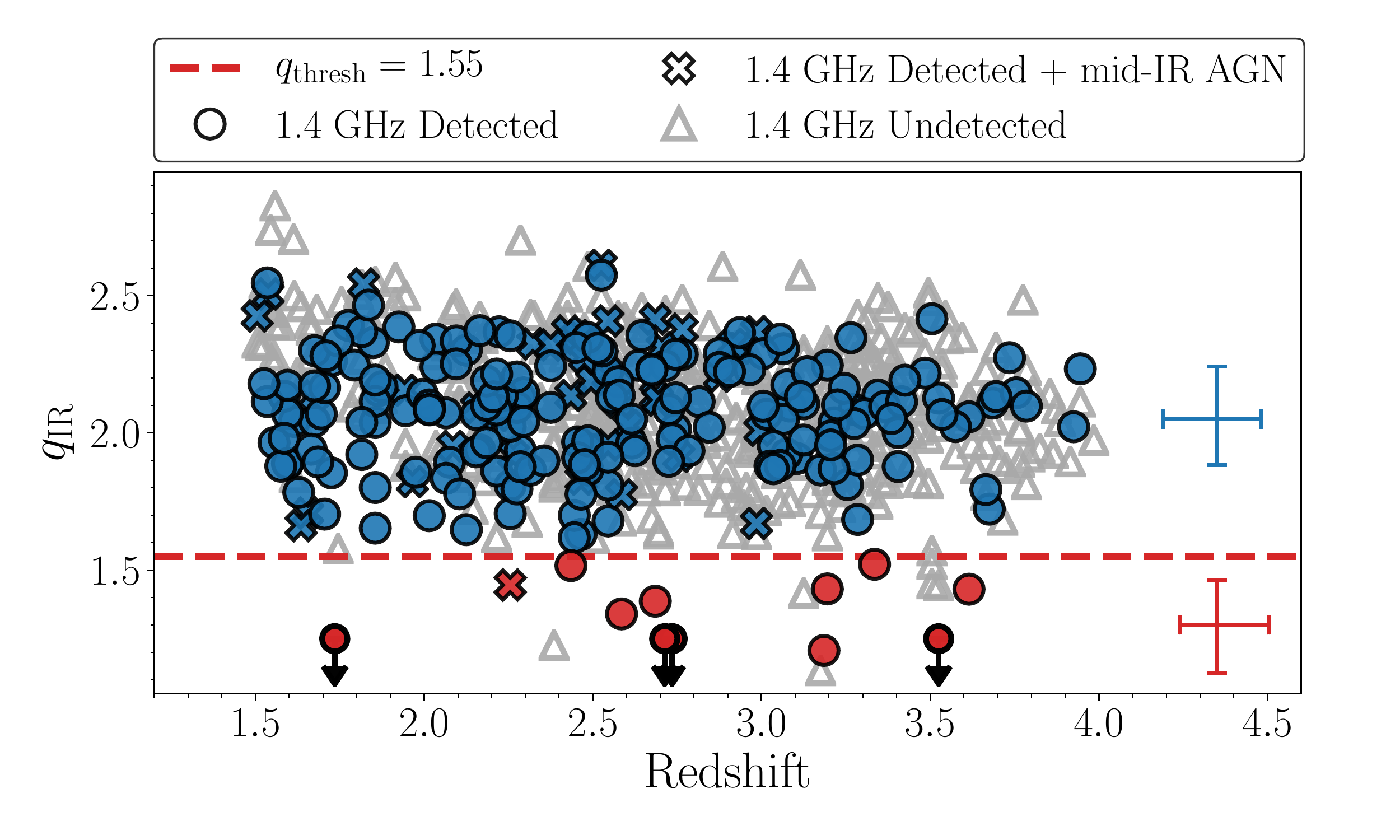}
    \caption{Distribution of $q_\text{IR}$ as a function of redshift for the AS2UDS SMGs within $1.5 \leq z \leq 4.0$. Galaxies with radio emission consistent with originating from star formation, defined as $q_\text{IR} > 1.55$ (red dashed line) are shown in blue, whereas radio-excess AGN are shown in red. The plotting limits are chosen to focus on the cloud of star-forming sources around $q_\text{IR} \sim 2.1$, which cuts off four radio-excess AGN within the range $q_\text{IR} = 0.35 - 0.95$. These are shown as red circles with downward pointing arrows. We additionally show two representative errorbars for the radio-detected star-forming and AGN populations. Lower limits on $q_\text{IR}$ are calculated using the corresponding upper limits on the SMG radio luminosity. Overall, AGN make up only $1.8 \pm 0.5\%$ of the SMG population, and hence the radio emission of the majority of SMGs is consistent with originating from star formation.}
    \label{fig:agn_definitions}
\end{figure*}

A subset of our SMG sample exhibits strong radio emission causing them to be substantially offset from the far-infrared/radio correlation for purely star-forming galaxies (Figure \ref{fig:agn_definitions}). This excess in radio power is attributed to additional emission from an active galactic nucleus in the centre of the galaxy, and hence forms a contaminant for studies of the far-infrared/radio correlation. As a result, such radio-excess AGN must be discarded from our sample, as it is not possible to disentangle the radio emission emanating from star formation or from the central AGN without resolving the radio emission, via e.g., very long baseline interferometry (e.g., \citealt{muxlow2005,muxlow2020,middelberg2013}). Typically, radio-excess AGN are seen to be hosted in red, passive galaxies \citep{smolcic2009b}. Nevertheless, about $1\%$ of local Ultra-Luminous Infra-Red Galaxies (ULIRGs) are also known to host such AGN \citep{condon1986,condon1991,yun1999}. Because our selection of SMGs does not involve their radio properties, it allows for an unbiased census of radio-excess AGN in high-redshift, strongly star-forming galaxies, as compared to previous radio-selected studies.

We identify AGN based on a fixed threshold of $q_\text{IR} \leq 1.55$, with sources below this threshold being defined as a radio-excess AGN (following e.g., \citealt{delmoro2013}). This value is chosen such that sources that are $\gtrsim5\times$ radio-brighter compared to the median (stacked) FIRRC for the AS2UDS sample, as derived in Section \ref{sec:results_qtir}, are identified as radio-excess AGN. Our threshold is similar to the value of $q_\text{IR} = 1.70$ adopted by \citet{thomson2014}, but takes into account that our typical $q_\text{IR}$ is slightly lower than that of their sample.

Upon adopting $q_\text{IR} = 1.55$ as our threshold, we find 12 radio-excess AGN within the full AS2UDS sample (Figure \ref{fig:agn_definitions}), corresponding to a surface density of $\sim12.5\pm3.6\,\text{deg}^{-2}$ at $S_{870} \gtrsim 4\,$mJy and $S_{1.4}\gtrsim30\,\mu\text{Jy\,beam}^{-1}$. Overall, $1.8 \pm 0.5 \%$ of SMGs therefore hosts a radio-excess AGN, similar to what is observed in local ULIRGs \citep{condon1986,condon1991,yun1999}. We have further investigated adopting other possible thresholds for identifying radio-excess sources, including using different cuts in $q_\text{IR}$, or adopting a redshift-dependent threshold in $q_\text{IR}$. The latter is commonly used for identifying AGN in radio-selected samples \citep{delhaize2017,calistrorivera2017}. However, we find that the far-infrared/radio correlation for AS2UDS is insensitive to the particular threshold we adopt, as the fraction of radio-excess AGN we identify among our sample is small regardless. As such, we proceed with a threshold of $q_\text{IR} = 1.55$.

\subsection{(A lack of) Redshift Evolution in the FIRRC}
\label{sec:results_qtir}

\begin{figure*}[!t]
    \centering
    \includegraphics[width=\textwidth]{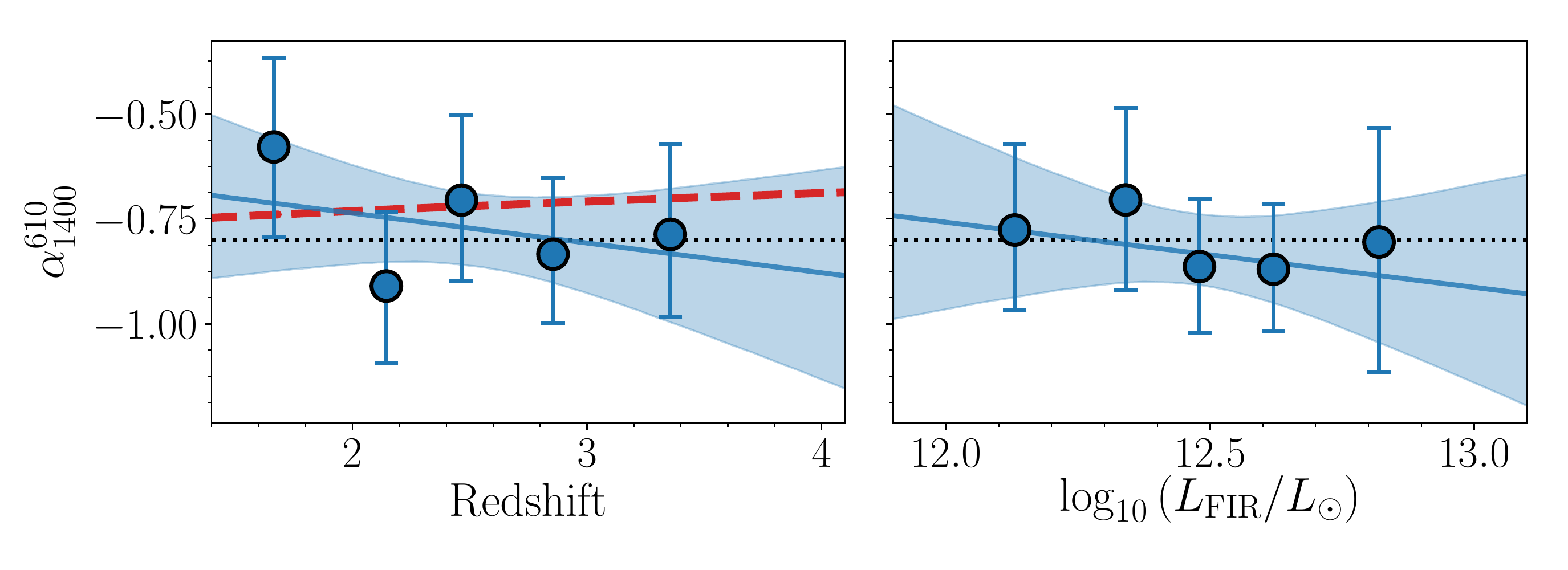}
    \caption{The spectral index between 610 MHz and 1.4 GHz for the full AS2UDS sample within $1.5 \leq z \leq 4.0$ (629 sources in total), computed for stacks in five bins in redshift (left) and FIR-luminosity (right). The expected redshift-evolution of the spectral index for an assumed synchrotron (free-free) spectral index of $\alpha = -0.85$ ($\alpha = -0.10$) and a thermal contribution of 10\% at rest-frame 1.4 GHz is shown via the red dashed line in the left panel (e.g., \citealt{condon1992}). In both panels a linear fit is shown via the blue line, with the uncertainty shown through the shaded region. The fits are consistent with no gradient in both redshift and far-infrared luminosity, and hence adopting a fixed $\alpha = -0.80$ (black dotted line) does not affect our calculation of the far-infrared/radio correlation.}
    \label{fig:alpha_vs_param}
\end{figure*}

In this section, we set out to constrain whether there is any redshift-evolution in the far-infrared/radio correlation for the AS2UDS sub-millimeter galaxies. In recent years, several studies have hinted at a decreasing value of $q_\text{IR}$ at increasing redshift. However, these studies have mainly been based on radio-selected samples (e.g., \citealt{delhaize2017,calistrorivera2017}) or optically selected samples (e.g., \citealt{magnelli2015}). \citet{thomson2014} carried out a study of the FIRRC based on a sub-millimeter selected sample from the ALESS survey. However, with a modest sample of $\sim70$ sources, \citet{thomson2014} were unable to distinguish between a redshift-independent far-infrared/radio correlation, or one where $q_\text{IR}$ decreases with redshift, as seen in radio-selected studies. With its tenfold increase in sample size, AS2UDS now provides a set of SMGs numerous enough to distinguish between these possible scenarios.

Before we proceed, we address one potential limitation of our analysis, which is the lack of available spectral indices for the majority of our radio sample. It has recently been suggested that a simple power-law approximation for the radio spectrum of highly star-forming galaxies may be insufficient, and that in fact radio spectra may exhibit a spectral break around a rest-frame frequency of $\sim5\,\text{GHz}$ \citep{tisanic2019,thomson2019}. For the full AS2UDS sample, where we probe rest-frame frequencies between $\nu_\text{rest} = 3.5- 7\,\text{GHz}$, any spectral steepening at high frequencies will affect the radio luminosities we calculate at rest-frame $1.4\,$GHz, which in turn will affect $q_\text{IR}$. A source at redshift $z$ with a true spectral index $\alpha$, for which a fixed value of $\alpha = -0.80$ was assumed, will have a calculated value of $q_\text{IR}$ which is off by $\Delta q_\text{IR} = -\left(0.80 + \alpha\right) \times \log_{10} \left( 1 + z \right)$, which amounts to approximately $0.2\,\text{dex}$ at $z=3$ for a spectral index equal to the 16$^\text{th}$ or $84^\text{th}$ percentiles of our $\alpha_{1400}^{610}-$distribution. Any systematic variations in the radio spectral index with redshift will therefore induce -- or potentially mask -- evolution in the FIRRC.

To assess the extent to which such variations might affect the far-infrared/radio correlation for AS2UDS, we stack the full SMG sample -- excluding radio AGN, but including sources undetected at 1.4\,GHz -- in five distinct redshift bins, in both the 610-MHz and 1.4-GHz maps. We additionally stack in $L_\text{FIR}$ and show the results in Figure \ref{fig:alpha_vs_param}. A linear fit through the data shows no evidence of spectral index evolution with either redshift or far-infrared luminosity, with a linear slope of $-0.07 \pm 0.16$ and $-0.15 \pm 0.42$ for the two parameters, respectively. The mean spectral indices are $\langle \alpha \rangle_z  = -0.76 \pm 0.07$ for the redshift bins, and $\langle \alpha \rangle_{L_\text{FIR}} = -0.80 \pm 0.04$ for the stacks in far-infrared luminosity. Both values are consistent with a typical spectral slope of $\alpha = -0.80$, as well as with each other, within the uncertainties. We further compare our values with the evolution expected in the spectral index when assuming an increasing contribution of free-free emission at high redshift, as a result of probing higher rest-frame frequencies for these galaxies. For this, we assume the simple model for star-forming galaxies from \citet{condon1992}, with a spectral index for synchrotron and free-free emission of, respectively, $\alpha_\text{synch} = -0.85$ and $\alpha_\text{FF} = -0.10$ (consistent with the values found by \citealt{niklas1997,murphy2011}). The expected flattening of the $610$-$1400\,\text{MHz}$ spectral index between $1.5 \leq z \leq 4.0$ is $\Delta\alpha \lesssim 0.1$, and we find no evidence for such modest evolution. This is fully consistent with \citet{thomson2019}, who in fact find a deficit in free-free emission for high-redshift SMGs. Overall, we find no significant variation in the $610-1400\,$MHz spectral index with either redshift or $L_\text{FIR}$, and we therefore conclude that the adopted radio spectral index is unlikely to be a driver of any trends in the AS2UDS far-infrared/radio correlation. \\

\begin{figure*}[!t]
    \centering
    \hspace*{-0.5cm}\includegraphics[width=1.0\textwidth]{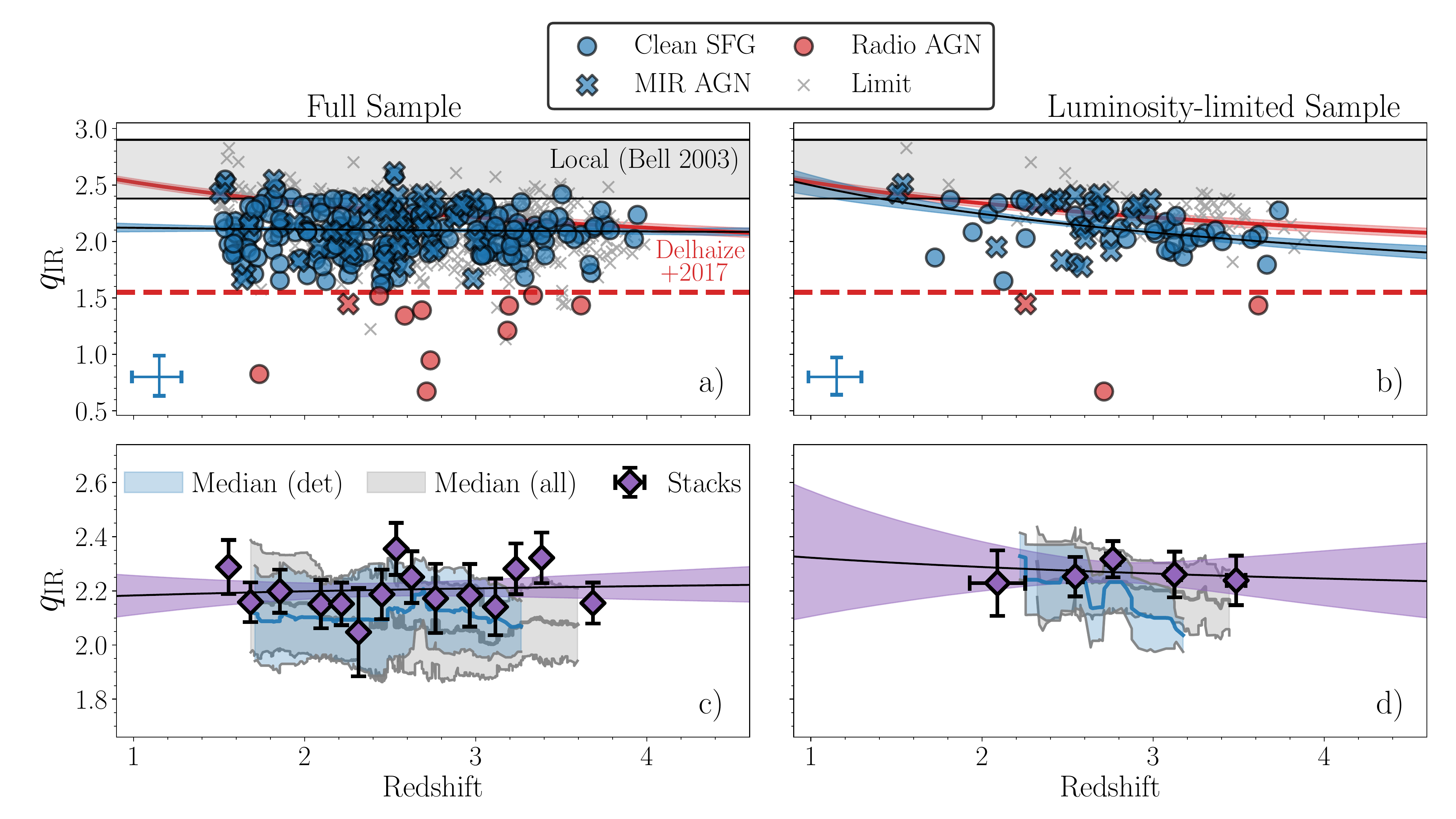}
    \caption{The far-infrared/radio correlation for AS2UDS as a function of redshift. \textbf{a)} the FIRRC for the full AS2UDS sample. The radio-detected star-forming sources are fitted by a power law of the form $q_\text{IR} \propto (1+z)^{\gamma}$. This fit and corresponding $1\sigma$ uncertainty are indicated via the black line and the blue, shaded region. The FIRRC for the full radio-detected AS2UDS sample shows no hint of redshift-evolution. For comparison, the local FIRRC and $1\sigma$ scatter from \citet{bell2003} is shown via the gray shaded band, and the evolving $q_\text{IR}$ from \citet{delhaize2017} is shown in red. A representative errorbar on $q_\text{IR}$ is shown in the bottom left corner, and star-forming sources and AGN are separated adopting a threshold of $q_\text{IR} = 1.55$ (dashed red line). \textbf{b)} the FIRRC for the radio-detected luminosity-limited AS2UDS sample. In contrast to the full sample, this subset \emph{does} show (artificial) evolution with redshift, as we select sources within a narrow range of $L_\text{FIR}$, but are only sensitive to the brightest radio sources at the high-redshift tail of AS2UDS. \textbf{c)} the FIRRC for the full AS2UDS sample, based on stacking in the 1.4-GHz radio map in 15 distinct redshift bins. The black line and purple shaded region show a power-law fit through these points, and its corresponding confidence interval. The blue (gray) shaded region shows the running median through the radio detections (detections + non-detections), where the spread indicates the median absolute deviation. The stacked full AS2UDS sample shows no hint of redshift-evolution. \textbf{d)} the stacked FIRRC for the luminosity-limited AS2UDS sample. In contrast to the radio-detections only, the stacked luminosity-limited sample shows no redshift-evolution in its far-infrared/radio correlation.}
    \label{fig:qTIR_detections}
\end{figure*}

We now proceed by investigating any potential redshift evolution in the far-infrared/radio correlation for sub-millimeter galaxies. In Figure \ref{fig:qTIR_detections} we show $q_\text{IR}$ as a function of redshift for the full AS2UDS sample and the luminosity-limited sample. In both cases, we fit a function of the form $q_\text{IR}(z) \propto (1+z)^\gamma$ to \emph{only} the SMGs detected at 1.4 GHz. As such, this sample is by construction biased towards radio-bright sources at higher redshift. For the full radio-detected AS2UDS sample, we find a lack of redshift-evolution, with a best fit solution of $\gamma_\text{full} = -0.01\pm0.03$. For the luminosity-limited sample, we do find an apparent evolution, and measure $\gamma_\text{lum} = -0.26 \pm 0.06$. However, this evolution is heavily driven by selection effects. While this sample is complete in far-infrared luminosity, the radio observations suffer from a positive $K$-correction, limiting the detection rate at high redshift. As a result, we are biased towards only the brightest radio sources at $z\gtrsim3$. For a fixed range in $L_\text{FIR}$ -- which the luminosity-limited sample is by construction -- radio-bright sources will have a low value of $q_\text{IR}$, and hence drive the average $q_\text{IR}$ down at higher redshift.

While the lack of redshift evolution for the full radio-detected AS2UDS sample -- which \emph{still} is biased -- is already interesting by itself, we need to address the radio-undetected population to get a proper census of any potential evolution of $q_\text{IR}$ across redshift. We do this by stacking the full and luminosity-limited samples in distinct redshift bins, having removed any radio AGN. We show $q_\text{IR}$ as a function of redshift for the stacked full and luminosity-limited samples in the bottom panels of Figure \ref{fig:qTIR_detections}. Neither sample shows any evidence of variation with redshift, with the full sample following a trend given by $\gamma_\text{full} = 0.02 \pm 0.06$, and the luminosity-limited sample having a best fit of $\gamma_\text{lum} = -0.02 \pm 0.16$. For reference, we additionally show the fifteen stacks and corresponding residuals of the full sample in Appendix \ref{app:stacking} (Figure \ref{fig:stacks}). We ensure the stacks are all of sufficient signal-to-noise ($\text{SNR}\gtrsim10$), such that reliable integrated flux measurements can be made, and any effects of noise boosting are minimal. As a result, the higher redshift bins contain a larger number of sources than the low-redshift ones, to compensate for the negative radio $K$-correction. We verified however, that the results are not affected by the method of binning, and simply adopting bins with an equal number of sources gives consistent results in all cases.

From the stacked results we further obtain an average value of $q_\text{IR}$ that, given our observed lack of redshift-evolution, is representative for sub-millimeter galaxies. For the full AS2UDS sample, we find a mean $q_\text{IR,full} = 2.20 \pm 0.03$, where the error represents the bootstrapped variation among the stacks. For the luminosity-limited sample, we find a similar value of $q_\text{IR,lum} = 2.26 \pm 0.02$, although across only five redshift bins. We further verify in Appendix \ref{app:stacking} that the expected systematic uncertainty on these values, as a result of our reliance on a stacking analysis, is small, and amounts to $\Delta q_\text{IR} \lesssim 0.05$. As the typical values of $q_\text{IR}$ for the full and luminosity-limited samples are consistent with one another, we will in the following investigate any possible trends between $q_\text{IR}$ and other physical parameters for the full AS2UDS sample, as its radio and far-infrared properties match those of the luminosity-limited subsample. Interestingly, this typical $q_\text{IR}$ for both samples is $\sim0.4\,$dex lower than the far-infrared/radio correlation observed locally \citep{bell2003}. We discuss this offset further in Section \ref{sec:discussion_evolution_SMG}.

\subsection{Correlations with Physical Properties}
\label{sec:results_correlations}

\begin{figure*}[t]
    \centering
    \includegraphics[width=1.0\textwidth]{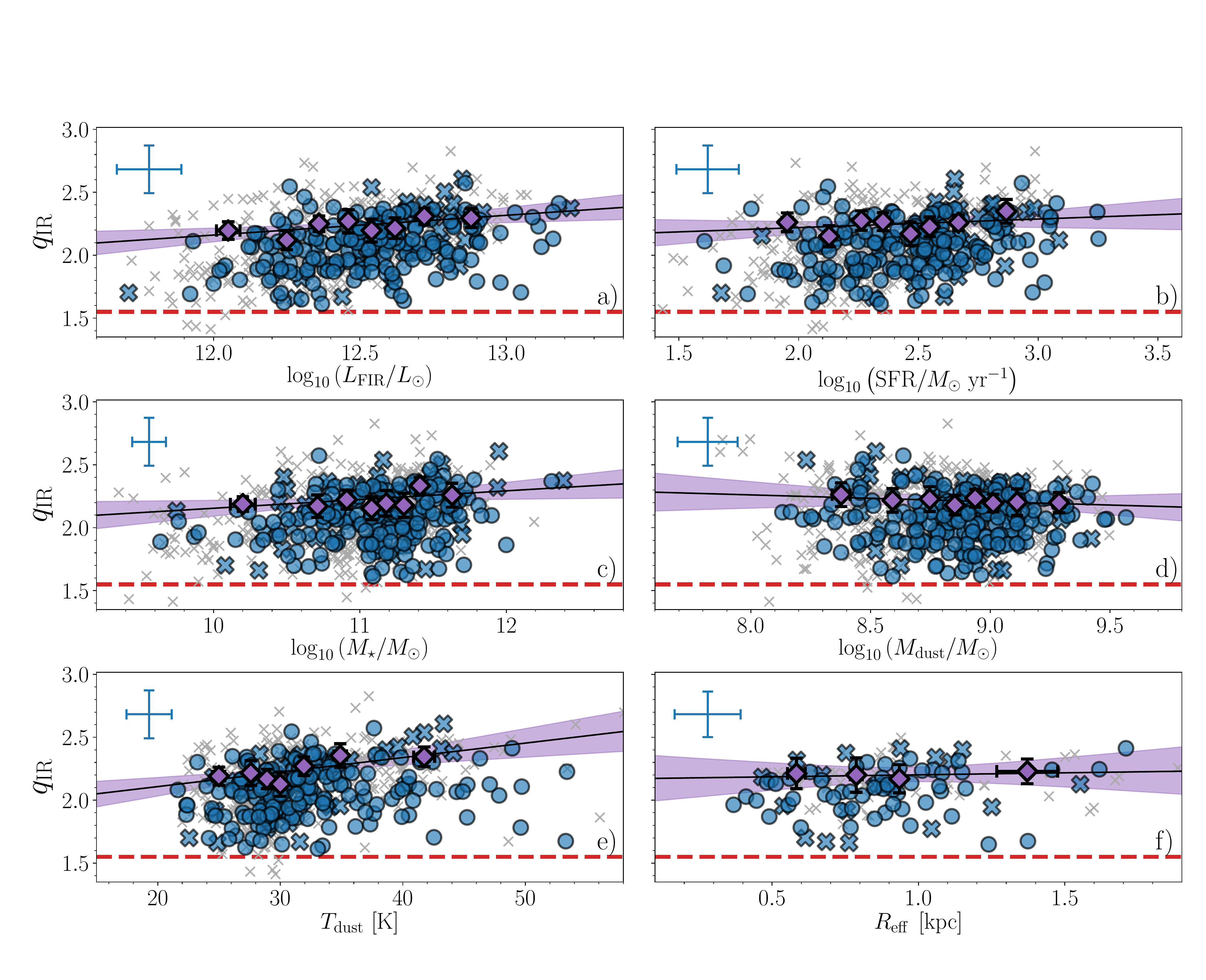}
    \caption{FIRRC parameter $q_\text{IR}$ as a function of several physical parameters for the full AS2UDS sample, after removal of radio-excess AGN. In all panels, we show radio-detected SMGs as blue circles (or blue crosses, when they show a mid-infrared AGN signature). A representative uncertainty for these is shown in the upper left corner of each panel. Lower limits on $q_\text{IR}$ are shown as gray crosses. The stacks are plotted as purple diamonds, with a linear fit to the stacks shown via the black line. The purple shaded region indicates the corresponding $16^\text{th}$-$84^\text{th}$ percentile confidence region on the fit. \textbf{a)} $q_\text{IR}$ as a function of far-infrared luminosity. \textbf{b)} $q_\text{IR}$ as a function of star-formation rate. \textbf{c)} $q_\text{IR}$ as a function of stellar mass. \textbf{d)} $q_\text{IR}$ as a function of dust mass. \textbf{e)} $q_\text{IR}$ as a function of dust temperature. \textbf{f)} $q_\text{IR}$ as a function of effective radius for sources with a robust submillimeter size measurement from \citet{gullberg2019}. None of the panels show any significant trends between $q_\text{IR}$ and the various physical parameters at a $\geq2\sigma$ level.}
    \label{fig:qTIR_correlations}
\end{figure*}

AS2UDS provides a large sample of SMGs for which \citet{dudzeviciute2019} have derived various physical properties via {\sc{magphys}}, such as stellar and dust masses, and star-formation rates. In this section, we investigate if there is any variation in $q_\text{IR}$ as a function of these parameters. In Figure \ref{fig:qTIR_correlations} we show $q_\text{IR}$ as a function of, respectively, $L_\text{FIR}$, $\text{SFR}$, $M_\star$, $M_\text{dust}$, $T_\text{dust}$ and effective observed-frame $870\,\mu$m-radius $R_\text{eff}$, the latter of which was calculated for a subset of submm-bright AS2UDS sources by \citet{gullberg2019}. In total, we have robust size measurements for 153 SMGs (70 are detected at 1.4 GHz). In order to assess the variation in $q_\text{IR}$ in an unbiased way, we perform a stacking analysis by dividing our SMG sample into distinct bins for the aforementioned physical parameters, after the removal of radio AGN. 

The first panel shows $q_\text{IR}$ as a function of infrared luminosity. While the radio-detected subset of AS2UDS follows a weak positive trend, any correlation disappears when stacking. A linear fit through the stacked datapoints indicates a slope of $\beta = 0.16 \pm 0.10$, consistent with no evolution. Similarly, no correlation between $q_\text{IR}$ and star-formation rate exists (slope of $\beta = 0.07 \pm 0.10$), which is expected since $L_\text{FIR}$ should be a good proxy for the star-formation rate of SMGs. 

Similarly, there does not appear to be any strong trend between $q_\text{IR}$ and stellar mass, with a linear fit through the stacks being consistent with a slope of zero ($\beta = 0.07 \pm 0.06$). Likewise, there is no evidence for any trends between $q_\text{IR}$ and either dust mass or temperature, with a slope of $\beta = -0.06 \pm 0.11$ and $\beta = (11 \pm 6)\times10^{-3}$, respectively. Since no trend with dust luminosity exists, which is a combination of $M_\text{dust}$ and $T_\text{dust}$, it is unsurprising that neither of these two parameters show any trend with $q_\text{IR}$ either. Finally, we show $q_\text{IR}$ as a function of $870\,\mu$m effective radius. As only a quarter of the full AS2UDS sample has measured submillimeter radii, we employ a smaller number of bins to obtain sufficient signal-to-noise in each stack. Nevertheless, we see no hint of a trend between $q_\text{IR}$ and $R_\text{eff}$, with a best-fitting linear slope of $\beta = 0.03 \pm 0.19$.

Overall, the AS2UDS SMGs do not appear to show any strong variation in $q_\text{IR}$ as a function of their physical properties. None of the six parameters explored show any hint of a correlation with $q_\text{IR}$ at a $2\sigma$ or greater level. This may be the result of the relatively small dynamic range spanned by the sample, or may in fact imply that the FIRRC constitutes an especially robust correlation, even at high star-formation rates and high redshift. We further discuss this in Section \ref{sec:discussion_evolution_SMG}.

\section{Discussion}
\label{sec:discussion}

\subsection{Previous Studies of the FIRRC}
\label{sec:comparison}
Neither the full AS2UDS sample, nor its radio-detected subset, show any evidence for evolution in their far-infrared/radio correlations. In this Section, we compare this lack of evolution with previous studies, including radio-based ones, which typically have large sample sizes, and SMG-based ones, having selection criteria that are more similar to ours.  

Recently, the FIRRC has been studied by \citet{delhaize2017} for the 3 GHz selected VLA-COSMOS sample \citep{smolcic2017a,smolcic2017b}. They utilize a sample of nearly 10,\,000 star-forming galaxies at a median redshift of $z\sim1.0$, and employ a survival analysis to attempt to account for non-detections at either radio or far-infrared wavelengths. They find statistically significant redshift-evolution of the FIRRC, with a slope of $\gamma_\text{D17}=-0.19\pm0.01$, out to $z\sim3$. \citet{molnar2018} extend this study by further tying in rest-frame ultraviolet morphological information for a subset of $\sim4700$ sources out to $z\sim1.5$. They split their sample into disk- and spheroid-dominated galaxies, and find that while the FIRRC for the latter shows significant redshift-evolution, similar to the study by \citet{delhaize2017}, the disk-dominated galaxies show minimal evolution, with a slope of $\gamma_\text{M18}=-0.037 \pm 0.012$. As radio AGN are typically found in red, bulge-dominated galaxies (e.g., \citealt{smolcic2009b}), this difference between the two samples is interpreted by \citet{molnar2018} as residual AGN contamination in spheroidal galaxies, and they argue the `true' FIRRC shows no evolution out to $z\sim1.5$.

The FIRRC was additionally studied at 1.4 GHz for a different radio sample by \citet{calistrorivera2017}, using Westerbork Synthesis Radio Telescope observations over the Bo\"{o}tes field. They include upper limits at both FIR- and radio wavelengths by using forced photometry, for a total of $\sim800$ sources. They too find siginificant redshift-evolution in the FIRRC at 1.4 GHz, out to $z\lesssim2.5$, with a slope of $\gamma_\text{CR17} = -0.15 \pm 0.03$, consistent with the aforementioned results from \citet{delhaize2017}. \\

Radio-selected samples, however, are by definition sensitive to radio-bright sources, and hence by construction select based on the combined radio luminosity from star-formation and AGN activity. Far-infrared-based surveys, in this regard, are mostly sensitive to emission solely from star-formation activity, as emission from a warm AGN torus is typically confined to mid-infrared wavelengths (e.g., \citealt{lyu2017,xu2020}). To substantiate this, we show in Appendix \ref{app:agn_fraction} that radio AGN are a factor of $\sim5$ more prevalent in radio-selected samples than in AS2UDS, at matched flux densities. As such, FIR-selected samples are expected to be substantially less contaminated by AGN.

For this reason, we now turn to two infrared-based studies of the far-infrared/radio correlation. We stress, however, that these typically have smaller sample sizes compared to radio-based surveys, but are less likely to suffer AGN contamination. \citet{ivison2010b} investigated the FIRRC out to $z\sim2$ using a \emph{Herschel} $250\,\mu$m selected sample over the GOODS-North field. They find modest evolution of $\gamma_\text{I10} = -0.26 \pm 0.07$ for a FIR-detected sample with $L_\text{FIR} = 10^{11} - 10^{12}\, L_\odot$, though their study is potentially affected by the large \emph{Herschel} point spread function and lack of high-resolution $250\,\mu$m identifications, complicating the association of radio counterparts to FIR-detections, and additionally complicating any stacking analyses. 

These problems were overcome by \citet{thomson2014}, who investigated the FIRRC for the ALESS $870\,\mu$m sample. Their sample selection is similar to that of AS2UDS, constituting an ALMA interferometic follow-up of submillimeter sources initially detected at the same wavelength in a single-dish survey \citep{karim2013,hodge2013}. The depth of both the ALESS and AS2UDS parent surveys and follow-up ALMA observations are roughly similar, as are the noise levels of the 1.4-GHz radio maps over the ECDFS and UDS fields, with the main difference being survey area and hence sample size. Therefore, ALESS forms the natural comparison sample to AS2UDS, and as such we compare the two surveys in additional detail.

\citet{thomson2014} individually detect 52 SMGs at 1.4 GHz, out of a parent SMG sample of 76 galaxies. We note that this parent sample excludes 21 SMGs that are optically faint, and hence had no reliable photometric redshift available (see also \citealt{simpson2014}). For the radio-detected subsample, \citet{thomson2014} find no evidence of redshift-evolution in the FIRRC, with a fitted slope of $\gamma_\text{T14} = -0.15 \pm 0.17$. Upon further including radio-undetected sources via a stacking analysis, they find a typical $q_\text{IR}$ across the full ALESS sample of $q_\text{IR} = 2.35 \pm 0.04$.\footnote{This is $\sim0.2\,$dex lower than was quoted in \citealt{thomson2014} (A. Thomson priv. comm.).} When limiting ourselves to the SMGs at $z \geq 1.5$ that do not exhibit a radio-excess signature, similar to our approach for AS2UDS, the ALESS sample shows a typical value of $q_\text{IR} = 2.33 \pm 0.04$. This is roughly similar to the typical value for AS2UDS of $q_\text{IR} = 2.20 \pm 0.03$. The small remaining difference of $\sim0.1\,$dex is likely the result of the slightly deeper SCUBA-2 map (typical RMS of $\sigma = 0.9\,\mu$Jy\,beam$^{-1}$, \citealt{geach2017,stach2019}) compared to the LESS parent survey for ALESS ($\sigma=1.2\,\mu$Jy\,beam$^{-1}$, \citealt{hodge2013}). Similarly, the AS2UDS ALMA observations are deeper than their ALESS counterparts. As a result, AS2UDS will identify somewhat infrared-fainter galaxies, which will decrease the typical $q_\text{IR}$ of the sample. We further note that \citet{thomson2019} study the far-infrared/radio correlation for a subset of 38 AS2UDS sources detected at both 1.4 and 6 GHz, for which they find a typical $q_\text{IR,T19} = 2.20 \pm 0.06$, consistent with the typical $q_\text{IR}$ we derive for the full AS2UDS sample. \\

Overall, while redshift-evolution of the far-infrared/radio correlation is near-unanimously found in radio surveys, evidence for such evolution when starting from infrared-selected samples is only weak. Both this observation and the aforementioned results from \citet{molnar2018} point towards unidentified radio AGN being the root cause of artificial evolution in the far-infrared/radio correlation in radio-selected surveys. However, we show in Appendix \ref{app:radio_firrc}, based on a combination of low-resolution radio and Very Large Baseline Interferometry (VLBI) observations in the COSMOS field, that this bias is insufficient. Summarizing, the VLBI data are predominantly sensitive to radio AGN -- however, the total radio emission in these high-resolution observations is not sufficient to explain the AGN contamination required in order to generate an evolving far-infrared/radio correlation, when compared to the total radio emission observed in the lower resolution Very Large Array radio observations.

\subsection{The FIRRC for SMGs}
\label{sec:discussion_evolution_SMG}

Observations of SMGs at high redshift have suggested that these systems are typically radio-bright compared to the local far-infrared/radio correlation (e.g., \citealt{kovacs2006,murphy2009a,magnelli2010}), though a clear demonstration of this offset has until now been complicated by the mostly small sample sizes employed, and their reliance on incomplete, radio-detected subsamples. For just the radio-detected AS2UDS SMGs, we find a typical $q_\text{IR} = 2.10 \pm 0.02$ (scatter $\sigma_q = 0.21$\,dex\footnote{This scatter is likely predominantly driven by the propagated measurement error on $q_\text{IR}$, which averages $0.18\,$dex.}), which is indeed substantially offset from the local correlation ($q_\text{IR} = 2.64$ with a scatter of 0.26\,dex, \citealt{bell2003}). However, this median value for AS2UDS is biased towards radio-bright sources as a result of selection. A truly representative value of $q_\text{IR}$ is obtained through our stacking analysis, which indicates a typical $q_\text{IR} = 2.20 \pm 0.03$. This implies that, even after correcting for selection effects, the FIRRC for our AS2UDS SMGs is offset from the local correlation for star-forming galaxies by $0.44\pm0.04\,$dex (a factor of $2.8\pm0.2$), while not showing any evidence for redshift-evolution between $1.5 \leq z \leq 4.0$ (a $3\sigma$ upper limit of $\leq 0.08\,$dex across this $\sim3\,$Gyr period). Consequently, this substantiates the finding of SMGs being radio-brighter relative to their FIR-luminosity compared to normal, star-forming galaxies found locally. 

The most straightforward explanation for this offset would be the contribution from an AGN to the observed radio emission. Based on the $0.4\,$dex offset from the local FIRRC, this requires the AGN to contribute $\sim70\%$ of the total radio luminosity. However, the small amount of scatter we observe around the correlation, as well as the low fraction of radio-excess AGN, requires substantial fine-tuning of AGN luminosities. VLBI observations further indicate a modest incidence of radio-AGN, with 3 out of 11 SMGs in the literature showing evidence for a compact core, indicative of an AGN (based on the combined samples of \citealt{biggs2010,momjian2010,chen2020}). These samples, in turn, explicitly target radio-bright SMGs, and the bright radio population is known to be dominated by radio-excess AGN (e.g., \citealt{condon1989}). As such, the incidence of dominant radio AGN in SMGs is likely to be a lot smaller than the $\sim30\%$ indicated by these VLBI studies.

Instead, both the offset in the FIRRC, as well as the lack of redshift-evolution for SMGs, are likely to be indicative of the different physics at play in normal, low-luminosity star-forming galaxies observed locally, and the much more active systems being studied at high redshift.

The calorimetric models of the far-infrared/radio correlation indeed make predictions for variations in the FIRRC as a function of star-formation surface density \citep{lacki2010a}, which may explain the difference between SMGs and the normal star-forming population. In addition, \citet{lacki2010b} model the behaviour of the FIRRC at high redshift, for galaxies with a variety of star-formation surface densities. With our large, homogeneous sample of SMGs, we can investigate the predictions of these models in detail. In the next section, we compare the far-infrared/radio correlation of the AS2UDS SMGs with that of normal star-forming galaxies. In Section \ref{sec:firrc_ulirgs}, we focus on the comparison with ULIRGs, thought to be the closest local analogs of $z\sim2$ dusty, star-forming galaxies.

\subsubsection{SMGs Compared to Normal Star-forming Galaxies}

Given that our low-frequency radio observations predominantly probe non-thermal synchrotron emission originating from relativistic electrons, we first discuss the far-infrared/radio correlation in terms of the various physical processes that compete for these electrons. In theory, the correlation is expected to break down at high redshift due to the increased inverse Compton losses of cosmic rays on the CMB (e.g., \citealt{murphy2009b,lacki2010b,schleicher2013}). Under the assumption that synchrotron and inverse Compton are the dominant processes of energy loss, a star-forming galaxy with a magnetic field of $B = 10\,\mu$G, as is typical for local, normal star-forming galaxies \citep{beck2013,tabatabaei2017}, is expected to show an increased $q_\text{IR}$ at $z=4$ compared to the local value of $\Delta q_\text{IR} \simeq 1.0\,$dex, as a result of the warmer CMB at high-redshift. Highly star-forming galaxies, however, are the most resilient to this, as their star-formation powered radiation fields are substantially stronger than the cosmic microwave background, even at moderate redshift. Under the assumption that our SMGs represent central starbursts with typical radius of $1\,$kpc (e.g., \citealt{gullberg2019}), the energy density $U_\text{rad}$ of their star-formation powered radiation field is still an order of magnitude higher than that of the CMB at $z=3$. The two energy densities are only expected to coincide at $z\sim6$, and due to the steep redshift-dependency of inverse Compton losses on the CMB ($U_\text{CMB}\propto(1+z)^4$, e.g., \citealt{murphy2009b}), such losses are negligible for the typical redshift range covered by sub-millimeter galaxies. As such, no evolution in the far-infrared/radio correlation is expected for the AS2UDS sample as a result of the warmer CMB at high redshift.

As we find the FIRRC for SMGs to constitute a particularly tight correlation, the relative radiative losses to synchrotron, inverse Compton and other potential sources of energy loss, such as ionization losses and bremsstrahlung (see e.g., \citealt{thompson2006,murphy2009b,lacki2010a}), have to be relatively constant across our sample (and hence, across redshift). This, too, is not surprising. We find no significant variation in $q_\text{IR}$ with a variety of physical parameters (Section \ref{sec:results_correlations}), neither for the individually radio-detected sources, nor for the stacks. \citet{dudzeviciute2019} further investigated any redshift-evolution for a variety of physical properties of the AS2UDS SMGs, and find only a strong increase in typical star-formation rates with increasing redshift. Further evolution in e.g., dust masses or gas fractions is only modest, and typically less than the differential evolution observed for the UDS field population. Overall, this paints the picture of SMGs as a fairly homogeneous galaxy population across redshift. Using a simple analytic model, \citet{dudzeviciute2019} explain the redshift distribution of SMGs as the combination of systems growing through a characteristic halo mass ($M_h \sim 4\times10^{12}\,M_\odot$) and acquiring a certain minimal gas fraction. If this threshold is associated with starburst activity, the SMG population might consist of physically similar galaxies, simply observed at different cosmic epochs. As radiative losses on the CMB remain negligible for our SMGs, as a result of the high star-formation powered radiation fields, the lack of redshift-evolution in the far-infrared/radio correlation of SMGs may simply be a consequence of their homogeneity. \\

This lack of evolution does however not explain the intrinsic offset of SMGs with respect to the local far-infrared/radio correlation. \citet{lacki2010a} argue that this offset is likely the result of the enhanced magnetic fields in SMGs, compared to those of the normal star-forming population. Neglecting, for now, other potential sources of cosmic ray energy loss besides inverse Compton, the fact that SMGs obey the far-infrared/radio correlation implies that $U_\text{B} / U_\text{rad} \gtrsim 1$ \citep{murphy2009b}, where $U_\text{B} = B^2 / 8\pi$. In other words, synchrotron emission must dominate the energy loss of cosmic rays, and the ratio of synchrotron to inverse Compton losses has to be relatively constant in general to explain the small scatter about the FIRRC. This, in turn, implies a minimum magnetic field strength for SMGs of $B_\text{min} \gtrsim 0.1 - 0.2\, \text{mG}$. Such magnetic fields are indeed expected for SMGs \citep{thompson2006,murphy2009b}, and are additionally in agreement with the $B-\text{SFR}$-relation deduced for local, normal star-forming galaxies by \citet{tabatabaei2017}, though we caution this requires an extrapolation across nearly two orders of magnitude in star-formation rate.

If ionization losses and bremsstrahlung are additionally expected to become important in highly star-forming galaxies, synchrotron emission has to be even stronger to maintain the far-infrared/radio correlation. In particular, enhanced synchrotron emission in SMGs is expected, as a result of their strong magnetic fields and what \citet{lacki2010a} call the `$\nu_\text{c}$-effect': a cosmic ray electron with an energy $E$ will predominantly emit synchrotron radiation at a frequency $\nu_\text{c}$, which is given by (e.g., \citealt{murphy2009b})

\begin{align}
    \left( \frac{\nu_c}{\text{GHz}} \right) = 1.3 \times \left( \frac{B}{0.1\,\text{mG}} \right)  \left( \frac{E}{\text{GeV}} \right)^2 \ .
    \label{eq:nu_c}
\end{align}

Hence, at a greater magnetic field strength, observations at a fixed frequency will probe lower-energy electrons. The distribution of injected electrons typically follows a power-law distribution in energy, $N(E) \propto E^{-p}$, where $p$ relates to the observed radio spectral index via $p = - (2\alpha - 1)$, in the absence of cooling. Typical values are $p > 2$, and in particular with $\alpha \approx -0.80$ we obtain $p\approx2.6$. This, in turn, implies that the lower typical energy of the electrons we probe is more than compensated for by them being substantially more numerous than their high-energy counterparts. This will, then, enhance the radio emission seen in SMGs. In particular, \citet{lacki2010a} propose that $q_\text{IR} \propto \left(1 - \frac{1}{2} p \right) \log_{10} B$. If we assume the offset of SMGs with respect to the local FIRRC is solely the effect of stronger magnetic fields in SMGs and the resulting $\nu_c$-effect, our observed $p=2.6$ implies SMGs have magnetic field strengths about 20 times larger than for typical local galaxies. As these generally have magnetic fields of $B\sim10\,\mu$G (e.g., \citealt{tabatabaei2017}), this implies that SMGs likely have magnetic fields of $B\sim0.2\,$mG, consistent with our previous minimum requirement on the field strength to maintain a linear far-infrared/radio correlation. While such magnetic fields are indeed strong compared to local, normal star-forming sources, they are smaller than the typical $\sim1\,$mG fields observed in local ULIRGs \citep{robishaw2008,mcbride2014}. Arp220, in particular, has an estimated magnetic field strength of $B\approx2\,$mG \citep{mcbride2015,yoasthull2016}. This difference is potentially due to ULIRGs being substantially more compact than SMGs, having similar levels of star formation in volumes of a few $100\,$pc \citep{solomon1997,downes1998}, instead of the $\sim\,$kpc scales that is typical for sub-millimeter galaxies \citep{simpson2015,hodge2016,gullberg2019}. We compare the far-infrared/radio correlation for SMGs and ULIRGs in more detail in Section \ref{sec:firrc_ulirgs}. \\

While seemingly satisfactory, simply enhancing the magnetic field of SMGs with respect to normal star-forming galaxies raises another issue, as was already noted by \citet{thompson2006}. The synchrotron cooling time is proportional to $B^{-3/2}$ \citep{murphy2009b}, and hence large magnetic field strengths imply very short synchrotron cooling times. This spectral ageing should in principle be observable in the synchrotron spectrum, manifesting itself as a spectral break. Such spectral features have indeed been claimed in the radio spectra of SMGs (e.g., \citealt{thomson2019}), at frequencies $\nu_\text{b} \gtrsim 5\,$GHz. For a single injection event of cosmic rays, subject to a magnetic field $B$, a spectral break arises at frequency $\nu_\text{b}$ after a time $\tau_\text{b}$, which is given by \citep{carilli1996}

\begin{align}
    \tau_\text{b} = 1.6\times\left(\frac{B}{0.1\,\text{mG}}\right)^{-3/2}\left(\frac{\nu_\text{b}}{\text{GHz}}\right)^{-1/2} \ \text{Myr}\ .
\end{align}

For a single, short burst of star formation, a spectral break at $5\,$GHz implies the synchrotron emission must have arisen within the last Myr, assuming $B=0.2\,$mG.\footnote{Given that we observe no deviations from a fixed spectral index of $\alpha = -0.80$ -- typical for uncooled synchrotron emission -- out to $z\approx4.0$, if a break exists, it is likely to lie at $\nu_\text{b}>5\,$GHz, which will further decrease $\tau_\text{b}$.} Additionally, for a top-hat star-formation history, modelled as a succession of single injection events following \citet{thomson2019}, the $610-1400$ MHz spectral index should have steepened to $\alpha \approx -1.2$ after only $20\,$Myr, with only minor differences when either a linearly rising, or exponentially declining star-formation history is assumed instead. Even accounting for the fact that synchrotron emission will lag the onset of the starburst by $\sim30\,$Myr \citep{bressan2002}, this is still inconsistent with the expected typical age of our SMGs of $\sim150\,$Myr, based on an analysis of depletion timescales \citep{dudzeviciute2019}.

At more realistic starburst timescales, the spectral break should manifest at much lower frequencies, and hence the $610 - 1400\,$MHz spectral index should be considerably steeper than the observed $\alpha \approx -0.8$. \citet{thompson2006} argue that, in the dense starburst environments, bremsstrahlung and ionization form additional sources of energy loss of cosmic ray electrons. Ionization losses, in particular, are most effective for low-energy cosmic rays, and hence will flatten the observed radio spectrum. For the cooling times for inverse Compton emission and ionization losses (respectively equations 4 and 10 in \citealt{murphy2009b}) to be equal, given a magnetic field of 0.1\,mG (1\,mG), requires an ISM density of $n_\text{ISM} \sim10^{3}\,\text{cm}^{-3}$ ($n_\text{ISM} \sim10^{2}\,\text{cm}^{-3}$). That is, for larger densities, ionization losses will dominate over inverse Compton cooling. Such densities are typical for the central regions of SMGs (e.g., \citealt{bothwell2013,rybak2019}), and hence the spectral steepening can be counteracted via ionization losses. In the models of \citet{thompson2006} and \citet{lacki2010b}, this indeed results in an expected $\alpha \approx -0.80$ at the rest-frame frequencies we probe for the AS2UDS SMGs,  which is consistent with our observations. \\

However, if ionization cooling is additionally important in sub-millimeter galaxies, it will compete with synchrotron emission for the available cosmic rays. In particular, increased ionization losses should work to reduce the observed synchrotron emission by a factor of $\sim2$ ($\Delta q_\text{IR} \approx +0.3\,$dex, \citealt{thompson2006}), which in turn partially compensates for the offset in the FIRRC as a result of the stronger magnetic fields in SMGs. To alleviate this tension, \citet{lacki2010a} suggest that the production of secondary cosmic ray electrons and positrons, resulting from proton-proton collisions in the high-density environment of a starburst galaxy, are generating additional synchrotron emission. Indeed, their models including the creation of secondary cosmic rays show a decrease of $\Delta q_\text{IR} \approx -0.4\,$dex at SMG-like gas surface densities, compared to models with only primary cosmic rays, which offsets the additional energy loss from bremsstrahlung and ionization losses. In particular, the creation of secondary cosmic rays should counteract strong spectral index gradients in galaxies hosting a central starburst, as these can be generated also outside the star-forming regions. While testing this at high-redshift is currently only possible in strongly gravitationally lensed galaxies (e.g., \citealt{thomson2015}), resolved multi-frequency observations of Arp220 between 150 MHz and 33 GHz indicate that cosmic ray electrons are required to be accelerated far outside the central regions in order to explain the spectral index maps \citep{varenius2016}, providing support to the importance of secondary cosmic rays (see also the discussion of multi-frequency source sizes in \citealt{thomson2019}). \\

Overall, our favored explanation for the lack of evolution in the far-infrared/radio correlation for SMGs, as well as its offset from the local value for normal star-forming galaxies, involves a fair amount of fine-tuning. Summarizing, it requires (i) strong magnetic fields ($B\gtrsim0.1-0.2\,$mG) to explain the offset in the FIRRC; (ii) significant ionization losses to counteract spectral ageing and flatten the observed radio spectra, and; (iii) secondary cosmic rays to compensate for this additional energy loss through ionization. This `conspiracy' indeed forms the basis for the models by \citet{lacki2010a,lacki2010b} in order to maintain a linear FIRRC across a wide range of far-infrared and radio luminosities. To test this scenario in more detail, ideally resolved radio- and far-infrared observations of a sizeable sample of SMGs are required. However, unresolved observations may be able to shed some light on the physical processes as well. The magnetic field strength of a galaxy likely depends on its star-formation activity, either through the gas surface or volume density and the Kennicutt-Schmidt relation \citep{lacki2010a,lacki2010b}, or directly via the observed star-formation rate \citep{tabatabaei2017}. If this correlation is continuous, one expects to see a negative correlation between $q_\text{IR}$ and star-formation rate, across a wide range from normal star-forming galaxies down to SMGs. 

If such a trend indeed exists, this will have a significant effect on studies of the far-infrared/radio correlation that are not uniformly sensitive to star formation across redshift. In particular, we argued in Section \ref{sec:comparison} and Appendix \ref{app:radio_firrc} that radio AGN alone cannot explain the observed redshift-evolution in radio-selected studies of the FIRRC. As such, it is probable that this evolution is instead the result of probing different galaxy populations locally and at high redshift. Unlike in the case of our far-infrared selected sample, selection at radio wavelengths is subject to a positive $K$-correction, such that at high redshift one is only sensitive to strongly star-forming galaxies (e.g., \citealt{condon1992}). In addition, the average high-redshift galaxy is more rapidly forming stars than a typical local star-forming galaxy (e.g., \citealt{speagle2014}). As such, radio-based studies will probe more `SMG-like' galaxies at high redshift, which implies that the average $q_\text{IR}$ probed should reflect the lower normalization for the far-infrared/radio correlation of SMGs. In turn, this should induce redshift-evolution in the far-infrared/radio correlation, similar to what is observed \citep{delhaize2017,calistrorivera2017}. The evolving $q_\text{IR}$ adopted in recent radio-based studies of the cosmic star-formation rate density (e.g., \citealt{novak2017,ocran2020}) should therefore be appropriate, as this quantity encompasses the modified conversion from radio emission to star-formation rate when considering different galaxy populations.

Testing this scenario, however, will require a star-forming sample with a fixed range of SFRs across redshift, where the star-formation rate is measured using a preferably dust-unbiased tracer independent of far-infrared and synchrotron emission. The most obvious candidates for this will be radio free-free emission and [\ion{C}{ii}]$\,158\,\mu$m emission, both of which suffer little from dust attenuation, and may provide an effective means of studying the far-infrared/radio correlation without requiring expensive, resolved far-infrared and radio observations.

\subsubsection{SMGs Compared to Local ULIRGs}
\label{sec:firrc_ulirgs}

The discussion in the previous section focused primarily on the difference in the far-infrared/radio correlation between SMGs and local, normal star-forming galaxies. However, ULIRGs potentially constitute the closest local analogues of $z\sim2.5$ sub-millimeter galaxies. They show far-infrared luminosities in excess of $10^{12}\,L_\odot$, and their typical magnetic field reaches mG strengths \citep{robishaw2008,mcbride2014,mcbride2015,yoasthull2016}, substantially larger than that of normal galaxies. However, ULIRGs \emph{do} fall onto the local far-infrared/radio correlation \citep{yun2001,farrah2003,jarvis2010,galvin2018}. If ULIRGs are indeed close analogs of SMGs, and magnetic fields are the primary driver of their different far-infrared/radio correlation, this raises the question why these seemingly similar galaxy populations are offset in $q_\text{IR}$ by $\sim0.4\,$dex.

While a detailed investigation of this offset is beyond the scope of this paper, we can afford to be more quantitative in a comparison between ULIRGs and SMGs. In the following, we will assume that all star formation is dust-obscured in both populations, and that the same physical processes for cosmic ray energy loss dominate. In particular, the strong magnetic fields and high densities present in both ULIRGs and SMGs indicate that the dominant processes are synchrotron, inverse Compton, ionization and bremsstrahlung. Their cooling times are given in \citet{murphy2009b}, and are reproduced here:

{\small
\begin{align}
\begin{split}
    \left(\frac{\tau_\text{syn}}{\text{yr}}\right) &\approx 4.4\times10^{4} \left( \frac{\nu}{\text{GHz}}\right)^{-1/2} \left( \frac{B}{\text{mG}}\right)^{-3/2} \\
    \left(\frac{\tau_\text{IC}}{\text{yr}}\right) &\approx 1.6\times10^{6}  \left( \frac{\nu}{\text{GHz}}\right)^{-1/2} \left( \frac{B}{\text{mG}}\right)^{1/2} \\
    &\times \left( \frac{L_\text{bol}}{10^{12}\,L_\odot}\right)^{-1}\left( \frac{R}{\text{kpc}} \right)^{2} \\
    \left(\frac{\tau_\text{ion}}{\text{yr}}\right) &\approx 1.1\times10^{9} \left( \frac{\nu}{\text{GHz}}\right)^{1/2} \left( \frac{B}{\text{mG}}\right)^{-1/2} \left( \frac{n_\text{ISM}}{\text{cm}^{-3}}\right)^{-1} \\
    &\times \left( \frac{3}{2}\left[ \ln\left(\frac{\nu}{\text{GHz}}\right) - \ln \left(\frac{B}{\text{mG}}\right) \right] + 39\right)^{-1} \\
    \left(\frac{\tau_\text{brem}}{\text{yr}}\right) &\approx 8.7\times10^{7} \left( \frac{n_\text{ISM}}{\text{cm}^{-3}}\right)^{-1} \ .
\label{eq:cooling}
\end{split}
\end{align}
}%

Here $L_\text{bol}$ is the bolometric luminosity of the galaxy, assumed to equal the $8-1000\,\mu$m luminosity, and $n_\text{ISM}$ is the particle number density of the interstellar medium. The fraction of cosmic ray energy that is emitted via synchrotron radiation can then be written as 

\begin{align}
    f_\text{syn} = \dfrac{1/\tau_\text{syn}}{\sum_{\text{proc},i} 1/\tau_i} \ ,
    \label{eq:fsyn}
\end{align}

where the sum iterates over all cooling timescales in Equation \ref{eq:cooling}. Crucially, $f_\text{syn}$ will depend on the frequency probed, as the various cooling timescales in Equation \ref{eq:cooling}, with the exception of bremsstrahlung, all contain a frequency-dependence. Synchrotron and inverse Compton losses are stronger at higher frequencies, as these probe more energetic particles (as per Equation \ref{eq:nu_c}). Ionization losses, on the other hand, are enhanced for less energetic particles, and will hence be weaker when probing higher rest-frame frequencies. Since we observe the AS2UDS SMGs at a fixed frequency of $\nu_\text{obs} = 1.4\,$GHz, the rest-frame frequency probed will be $\nu_\text{rest} = 1.4\times(1+z)\,$GHz. As was noted in \citet{lacki2010b}, this implies that the observed radio luminosity $L_{\nu_\text{rest}}$ will be proportional to $f_\text{syn}(\nu_\text{rest})$. An observer will then $K$-correct $L_{\nu_\text{rest}}$ to rest-frame 1.4 GHz with a given spectral index. However, since our SMG sample spans a range of redshifts, yet is observed at a fixed frequency, a range of $f_\text{syn}(\nu_\text{rest})$ is probed, and as such $f_\text{syn}$ will be a function of redshift. Crucially, this implies that $q_\text{IR}$ too will vary with redshift as

\begin{align}
    q_\text{IR} = q_0 - \log_{10}\left( f_\text{syn}(z) \right) \ ,
    \label{eq:qIR_fsyn}
\end{align}

for some a priori unknown normalization $q_0$, under the assumption that the SMG population itself does not evolve significantly with redshift. If the same physical processes are at play in shaping the far-infrared/radio correlation for SMGs and ULIRGs, we can use the observed normalization of the local FIRRC for the latter to model the expected evolution in the correlation for SMGs. In particular, we may write

\begin{align}
    q_\text{IR}(z) = 2.64 + \log_{10}\left( \frac{ f_\text{syn}^\text{ULIRG} }{ f_\text{syn}^\text{SMG}(z) } \right) \ ,
    \label{eq:qIR_ULIRGs}
\end{align}

where the normalization of $q_0=2.64$ from \citet{bell2003} was adopted for ULIRGs. We note this is consistent with $q_\text{IR} = 2.70 \pm 0.06$ from \citet{farrah2003} and $q_\text{IR} = 2.64 \pm 0.01$ measured by \citet{yun2001}.\footnote{For \citet{yun2001}, we convert FIR-luminosities from $42.5-122.5\,\mu$m to the $8-1000\,\mu$m range used in this work by adding $0.30\,$dex, following \citet{bell2003,delhaize2017}.}

Parameter $f_\text{syn}$ is fully determined, under our simplifying assumptions, by the magnetic field strength, ISM density, physical size, far-infrared luminosity and redshift of any given source. As such, we adopt a set of standard values for these parameters for both ULIRGs and SMGs, as tabulated in Table \ref{tab:parameters}. We note that the magnetic field strength and particle densities are not particularly well-constrained in either, and as such this `benchmark' model comes with inherent uncertainties. Nevertheless, for simplicity we adopt equal magnetic field strengths of $B=1.0\,$mG for SMGs and ULIRGs, which is consistent with the lower limit of $B\gtrsim0.2\,$mG we determined for SMGs in order to maintain a linear far-infrared/radio correlation in the previous Section. We further adopt far-infrared luminosities typical for the \citet{farrah2003} and AS2UDS samples for ULIRGs and SMGs, respectively. In addition, we adopt a typical effective radius for the dust emission in SMGs of $1.0\,$kpc (e.g., \citealt{gullberg2019}), and adopt $250\,$pc for ULIRGs (e.g., \citealt{solomon1997,downes1998}). The far-infrared/radio correlation of SMGs, however, is not particularly sensitive to the adopted physical size, since it only affects inverse Compton losses, as per Equation \ref{eq:cooling}. Cosmic rays in the more compact ULIRGs, however, lose a substantially larger fraction of their energy via inverse Compton cooling, as ULIRGs have far-infrared luminosities comparable to those of SMGs, yet condensed into a smaller volume. Given their increased compactness, we additionally adopt a larger typical ISM density in ULIRGs than in SMGs, although we stress both are inherently uncertain. Given these limitations, we do not aim to make strong predictions on the physical conditions in either ULIRGs or SMGs, but instead use this simplified model to explain global trends in their far-infrared/radio correlation.

\begin{deluxetable}{llll}
    \centering
	\tabletypesize{\footnotesize}
	\tablecolumns{4}
	\tablewidth{0.48\textwidth}
	\tablecaption{Benchmark Models for ULIRGs and SMGs}
	
	\tablehead{
		\colhead{\textbf{Parameter}} &
		\colhead{\textbf{Unit}} &
		\colhead{\textbf{ULIRG}} &
		\colhead{\textbf{SMG}}
	}
	
	\startdata
	
	$\log_{10} L_\text{IR}$ & $L_\odot$ & $12.0$ & $12.5$  \\
	$R_\text{eff}$ & kpc & 0.25 & 1.0 \\
	$B$ & mG & 1.0 & 1.0 \\
	$ n_\text{ISM}$ & $\text{cm}^{-3}$ & $2\times10^{3}$ & $1\times10^{3}$ 
	\enddata
	\label{tab:parameters}
\end{deluxetable}

\begin{figure*}[!t]
    \centering
    \includegraphics[width=1.0\textwidth]{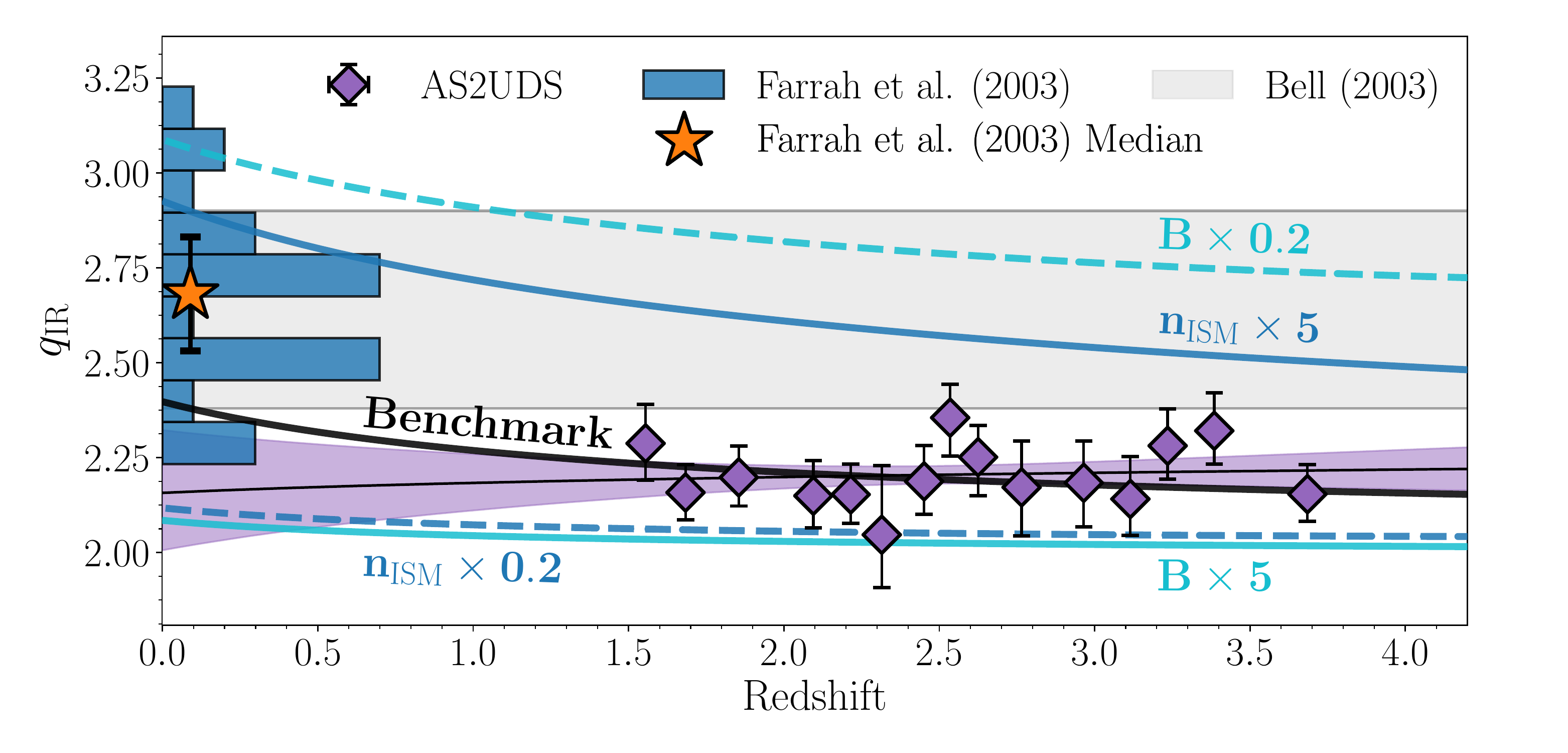}
    \caption{A comparison of the far-infrared/radio correlation for the AS2UDS SMGs with a heterogeneous mix of local, star-forming galaxies \citep{bell2003} and local ULIRGs \citep{farrah2003}. The AS2UDS points and fit from Figure \ref{fig:qTIR_detections}c are shown in purple. In addition, five evolutionary toy models for the FIRRC are overlaid. The benchmark model adopts the same magnetic field strength in SMGs and ULIRGs, but assumes the latter are both more compact, and more dense. Four variations are also shown, adopting different magnetic field strengths and densities for SMGs relative to the benchmark model. The increased compactness of ULIRGs compared to $z\sim2.5$ dusty star-forming galaxies results in higher ionization and inverse Compton losses, and hence in a subdominant contribution from synchrotron emission. This, in turn gives rise to an increased $q_\text{IR}$.}
    \label{fig:qIR_vs_local}
\end{figure*}

We plot the expected $q_\text{IR}(z)$ for SMGs, normalized to the local far-infrared/radio correlation for ULIRGs, in Figure \ref{fig:qIR_vs_local}. In addition, we indicate how this trend is affected by an increase/decrease in $B$ or $n_\text{ISM}$ by a factor of five. It is clear that our benchmark model recovers the correct normalization of the FIRRC for SMGs. In addition, the frequency-dependence of the cooling times induces a slight redshift-dependency, which, if fitted by $q_\text{IR} \propto (1+z)^\gamma$, implies that $\gamma \approx -0.05$, which is marginally consistent with our results. Note that an extrapolation of this model to $z=0$ results in a typical $q_\text{IR} \approx 2.4$, which is below the local far-infrared/radio correlation. This should however not be interpreted as the predicted normalization for local ULIRGs. Instead, this is the typical $q_\text{IR}$ an SMG would have when a rest-frame frequency of 1.4\,GHz is probed directly.

The variations on the benchmark model in Figure \ref{fig:qIR_vs_local} indicate that $q_\text{IR}$ is substantially affected by changes in the density or magnetic field strength. Increasing the latter naturally increases the relative contribution of synchrotron emission to the overall cosmic ray energy loss, and as such decreases $q_\text{IR}$ (Equation \ref{eq:qIR_fsyn}). Decreasing the density has an analogous effect, as the relative contribution of ionization losses is diminished, and hence $f_\text{syn}$ is enhanced. We emphasize that our benchmark model is simply one of a family of models with the correct behaviour, reproducing the normalization and lack of strong redshift-evolution in the FIRRC for SMGs. However, all such models require that synchrotron emission is dominant in SMGs, with a subdominant contribution from ionization losses. A large value of $f_\text{syn} \gtrsim 0.6$ is additionally required to obtain a relatively flat slope in the $q_\text{IR}$-redshift plane. In ULIRGs, however, synchrotron emission is subdominant ($f_\text{syn} < 0.5$), and instead substantial contributions from ionization and inverse Compton losses ensure that their far-infrared/radio correlation is offset from that of SMGs, and is consistent with that of typical star-forming galaxies observed locally.

The interpretation that synchrotron emission is subdominant in ULIRGs implies that these should have flatter radio spectral indices compared to SMGs due to the increased importance of ionization losses, at comparable rest-frame frequencies. This is in agreement with results in the literature, which indicate that ULIRGs typically show a $1.4-5\,$GHz spectral index of $\alpha \approx -0.50$ to $-0.60$ \citep{clemens2008,leroy2011,galvin2016,klein2018}. Our $610\,$MHz$-1.4\,$GHz observations of $z\sim2-3$ SMGs probe identical rest-frame frequencies as the higher frequency data for local ULIRGs, but these instead show a steeper $\alpha \approx -0.80$. While \citet{clemens2008} interpret this flattening in ULIRGs as the result of increased free-free absorption, they note that ionization losses will have a similar effect on the spectral index. Since no such flattening is observed in SMGs at identical rest-frame frequencies, we prefer the latter interpretation. While we rely mostly on stacked spectral indices in this work, the combination of matched-depth 610\,MHz and 1.4\,GHz observations of SMGs will provide a suitable means to investigate this in additional detail. 

The fact that synchrotron emission is subdominant in ULIRGs further implies that the intrinsic scatter about their far-infrared/radio correlation is likely to be enhanced, as small deviations in e.g., the magnetic field strength or density will have a relatively large impact on the observed $f_\text{syn}$, and in turn on the $q_\text{IR}$ of individual sources. Such an increase in the scatter about the FIRRC at high far-infrared luminosities has been widely observed \citep{helou1985,condon1991c,yun2001,bressan2002,bell2003}, and may hence be related to the subdominance of synchrotron emission in ULIRGs. However, \citet{bressan2002} instead interpret this increased scatter as being due to a timescale effect, as far-infrared emission arises on a shorter timescale than radio emission after the onset of star formation. If ULIRGs constitute strong yet relatively short-lived starbursts, the scatter about their far-infrared/radio correlation will be enhanced compared to that of more continuously star-forming systems. In order to test these interpretations, an indicator of the starburst age is required. Multi-frequency radio observations may provide a way forward here too, as spectral indices form a proxy for starburst age. In combination with other age-indicators, such as star-formation rate tracers that probe different timescales, it is possible to begin reconstructing the recent star-formation history observationally. In addition, with multi-frequency radio observations one can eliminate the inherent uncertainties arising from the intrinsic scatter expected about the far-infrared/radio correlation when a fixed spectral index is assumed.

\section{Conclusions}
\label{sec:conclusion}
We have presented a study of the far-infrared/radio correlation for AS2UDS \citep{stach2019,dudzeviciute2019} -- a homogeneously selected sample of $706$ sub-millimeter galaxies identified through ALMA band 7 follow-up of SCUBA-2 $850\,\mu$m observations of the UKIDSS/UDS field. Through combining these sub-millimeter data with deep ($\sigma_\text{RMS}\sim7\,\mu$Jy\,beam$^{-1}$) Very Large Array radio observations at $1.4\,$GHz and available 610\,MHz coverage from the GMRT, we study their joint far-infrared and radio properties within the redshift range $1.5 \leq z \leq 4.0$, where our selection is uniform in terms of dust mass or far-infrared luminosity. 

We address the incompleteness in the radio observations through a stacking analysis, and find a typical far-infrared/radio correlation parameter $q_\text{IR} = 2.20 \pm 0.03$ for sub-millimeter galaxies, which is $\sim0.4\,$ dex lower than the local value for a heterogeneous mix of star-forming galaxies \citep{bell2003} and ULIRGs \citep{yun2001,farrah2003}. This value of $q_\text{IR}$ further shows no evidence of evolution between $1.5 \leq z \leq 4.0$, which likely illustrates that SMGs are a physically homogeneous population of galaxies across redshift, for which no strong redshift-evolution is expected. 

This offset for SMGs with respect to the local far-infrared/radio correlation is unlikely to be the result of residual contamination by radio AGN, which is more likely to affect radio-selected samples. Instead, we interpret the offset with respect to the far-infrared/radio correlation of normal galaxies through strong magnetic fields in SMGs ($B \gtrsim 0.2\,$mG), combined with the production of secondary cosmic rays, both of which serve to increase the radio output of a galaxy for a given star-formation rate \citep{lacki2010a,lacki2010b}. Combined high-resolution radio and far-infrared observations of a large sample of sub-millimeter galaxies will provide a robust way to test this interpretation in the future. We additionally model the offset in the FIRRC between SMGs and local ULIRGs, under the assumption that the same physical processes are at play in either. In particular, a model wherein ULIRGs are denser and more compact than SMGs can fully explain the observed offset, as well as the lack of evolution in the far-infrared/radio correlation of SMGs. A prediction of this interpretation is that SMGs have steeper GHz radio spectral indices than ULIRGs, and a reduced scatter about their far-infrared/radio correlation. We argue that matched-depth, multi-frequency radio observations of SMGs are crucial in order to test both of these predictions.

\section*{Acknowledgements}
We wish to thank the anonymous referee for their comments and suggestions which have improved this work. HSBA, JAH and DvdV acknowledge support of the VIDI research programme with project number 639.042.611, which is (partly) financed by the Netherlands Organization for Scientific Research (NWO). IS, UD, AMS and SS acknowledge support from STFC (ST/T000244/1). EdC gratefully acknowledges the Australian Research Council as the recipient of a Future Fellowship (project FT150100079). CCC acknowledges support from the Ministry of Science and Technology of Taiwan (MOST 109-2112-M-001-016-MY3). JLW acknowledges support from an STFC Ernest Rutherford Fellowship (ST/P004784/1 and ST/P004784/2). The National Radio Astronomy Observatory is a facility of the National Science Foundation operated under cooperative agreement by Associated Universities, Inc. The ALMA data used in this paper were obtained under programs ADS/JAO.ALMA\#2012.1.00090.S, \#2015.1.01528.S and \#2016.1.00434.S. ALMA is a partnership of ESO (representing its member states), NSF (USA) and NINS (Japan), together with NRC (Canada) and NSC and ASIAA (Taiwan), in cooperation with the Republic of Chile. The Joint ALMA Observatory is operated by ESO, AUI/NRAO, and NAOJ.

\appendix

\section{Stacking}
\label{app:stacking}
Throughout this work, we employ a stacking technique in order to incorporate the radio-undetected population. While in principle a straightforward procedure, in practice stacking involves a variety of choices. In particular, the first choice one must make, is whether to adopt either the mean or median when stacking. In this work, we adopt the latter, as the median is considerably less affected by outliers in the underlying distribution of flux densities (e.g., \citealt{white2007}). In addition, \citet{condon2013} show that the mean is likely to be significantly affected by sources close to the survey detection limit, and as such may be less representative of the full underlying population. We have verified, however, that our results are unchanged when adopting a clipped-mean stacking procedure, whereby SMGs with nearby bright radio sources are omitted from the stacking. While this agreement is encouraging, we still prefer the median as it does not require a (somewhat arbitrary) rejection of sources when stacking, in addition to the reasons elaborated above.

\begin{figure}[!t]
    \centering
    \hspace*{-0.3cm}\includegraphics[width=0.5\textwidth]{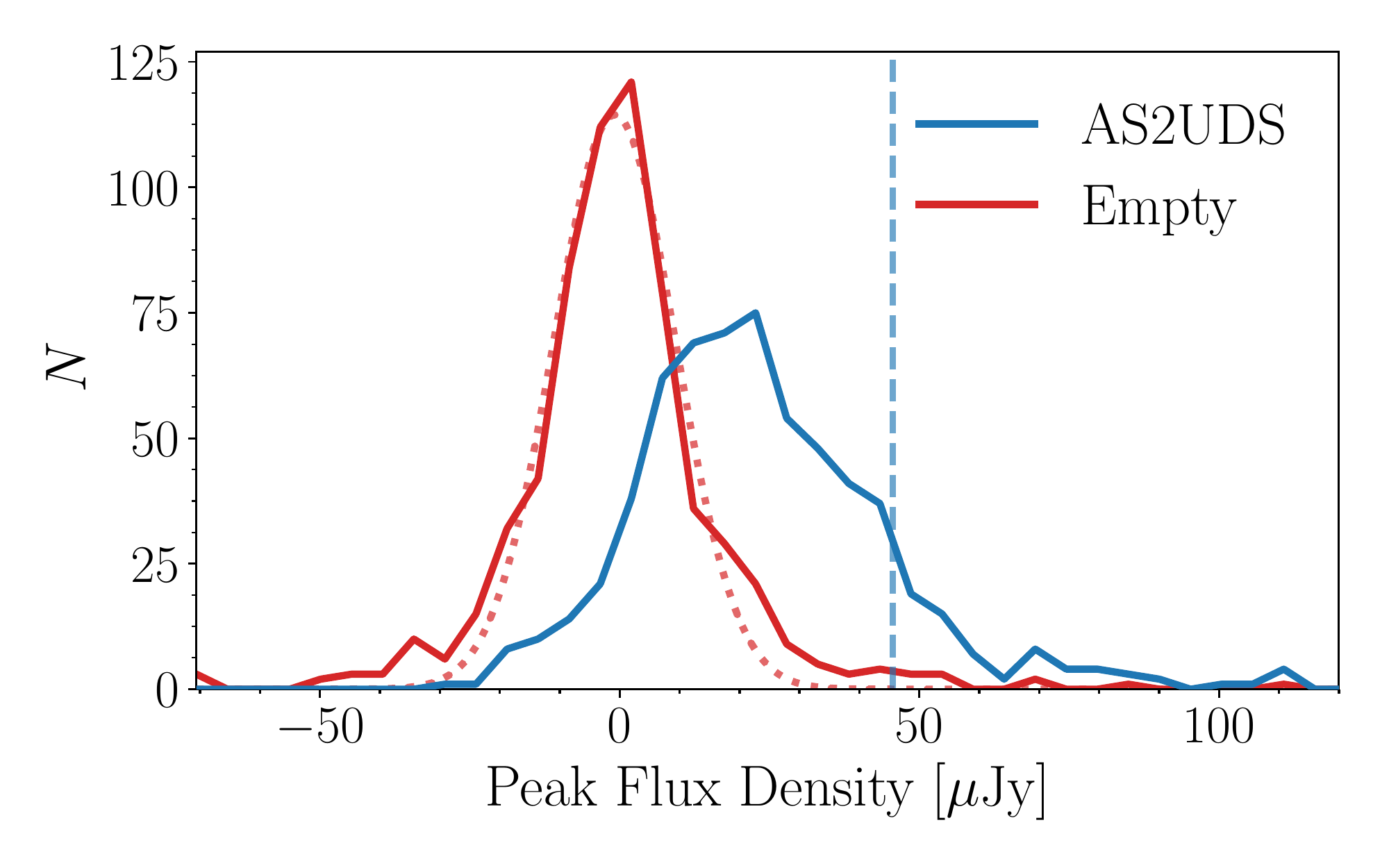}
    \caption{Distribution of peak-pixel flux densities for the AS2UDS SMGs, and empty background regions. The red, dotted line shows a Gaussian fit to the latter, with a mean of approximately zero, and the vertical dashed line indicates the detection limit ($4\sigma$) at 1.4\,GHz. The peak-pixel values for AS2UDS show a clear positive excess, indicative of a large number of sources below the detection limit.}
    \label{fig:stack_histograms}
\end{figure}

A second choice one must make is how to measure stacked flux densities. In radio astronomy, typically the peak flux density is adopted for a source that is unresolved, while the integrated flux density is utilized for extended sources. Ideally, for a set of unresolved radio sources, one would simply take the pixel value at the location of the sub-millimeter galaxy, which is equivalent to adopting the peak flux density when all sources are perfectly aligned, in the absence of noise. We show the distribution of peak pixel values in the 1.4\,GHz map for the AS2UDS SMGs in Figure \ref{fig:stack_histograms}, and further verify that the fluxes in the off-source stacks used for determining the background and RMS-noise level are Gaussian and centered around zero. The distribution of fluxes at the AS2UDS source positions are consistent with a power-law at high flux densities, and are dominated by the local Gaussian noise in the mosaic at flux densities near the typical RMS in the map.

However, these peak-pixel flux densities are likely to underestimate the true flux density of our stacks, as there might be spatial offsets between the radio and far-infrared emission -- either real or as a result of the local noise in the maps. In addition, the sources are not expected to be perfectly unresolved -- for instance, at a resolution of $1\farcs8\times1\farcs6$ at 1.4\,GHz, nearly half of the radio-detected AS2UDS SMGs are (marginally) resolved at the $3\sigma$ level, based on their deconvolved source sizes (following \citealt{thomson2019}). As such, integrated flux densities are likely to be more appropriate for the AS2UDS stacks at 1.4\,GHz. This is further substantiated by the fact that the integrated flux densities for the stacks of the full AS2UDS sample in Figure \ref{fig:qTIR_detections} exceed the peak flux density by a typical factor of $S_\text{int} / S_\text{peak} = 1.64 \pm 0.08$. However, in order to determine integrated flux densities, one typically fits Gaussians to the stacked detections, which in turn may be biased when the stacks are at low signal-to-noise. We investigate this via a mock-stacking procedure, whereby we insert faint (i.e., including flux densities well below the detection limit) mock sources into the UDS 1.4\,GHz map, stack them, and compare both the inserted and recovered flux densities, as well as the recovered peak and integrated values. We adopt a power-law distribution in inserted mock source flux densities with a slope of $-2$, which is typical for radio number counts for star-forming sources in the sub-mJy regime (e.g., \citealt{prandoni2018}), and is additionally consistent with the distribution of peak pixel fluxes shown in Figure \ref{fig:stack_histograms}. The number of sources sampled from this distribution was further varied between $N=30-100$ in order to match the typical number of sources stacked in this work. We then compare the results of two runs: in the first, mock sources are inserted as unresolved, while in the second $40\%$ of mock sources are slightly resolved, randomly being assigned a Gaussian size between $1.2-1.6\times$ the beam size. In addition, in order to mimic a true survey, the catalogued mock source positions -- those which are used for stacking -- are slightly different from the true mock source positions, as we randomly draw an offset in both right ascension and declination from a Gaussian distribution with zero mean and standard deviation of $\sigma=0.30''$. This positional offset is consistent with the observed distribution of separations between the far-infrared and radio positions of the AS2UDS sources with radio counterparts at 1.4\,GHz.

\begin{figure*}[!t]
    \centering
    \includegraphics[width=\textwidth]{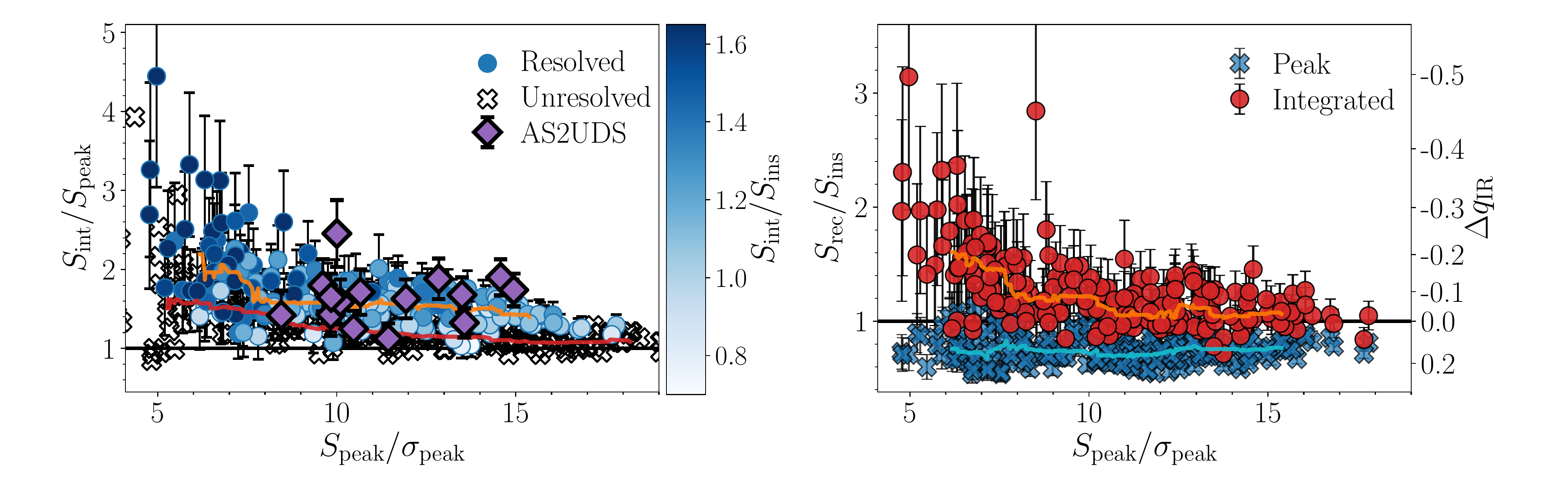}
    \caption{\textbf{Left:} The ratio of the recovered integrated and peak flux densities of the stacked mock sources, as a function of peak SNR. The data are colored by the magnitude of flux boosting. Black crosses correspond to stacking of unresolved mock sources only, while the circles include resolved sources, as well as small, random positional offsets between the true and catalogued source centers. The running median through the circles (crosses) is shown via the orange (red) line. The stacked AS2UDS SMGs (purple diamonds) show a substantially larger typical integrated-to-peak flux density ratio than the unresolved mock sources. \textbf{Right:} The ratio of recovered/inserted flux density versus recovered peak SNR. The orange (blue) line indicates the running median through the points representing integrated (peak) flux density measurements, and the right ordinate axis shows the propagated offset in $q_\text{IR}$ based on the difference in the true and recovered flux densities. Overall, the integrated value constitutes a better estimate of the true flux density of the stacks at $\text{SNR} \gtrsim 10$.}
    \label{fig:mockstacking}
\end{figure*}

We compare the two mock-stacking runs in Figure \ref{fig:mockstacking}, where we quantify the difference between recovered integrated and peak flux densities, as well as the level of ``flux boosting'', that is, the ratio of the recovered and inserted flux density. For unresolved mock sources, the integrated-to-peak ratio is typically slightly larger than unity, indicating that at a modest $\text{SNR}\lesssim10$, integrated flux densities may be slightly overestimated. However, at the typical $\text{SNR}\sim8.5-15$ we obtain for the AS2UDS stacks, $S_\text{int}/S_\text{peak}=1.18\pm0.02$ for the unresolved mock sources, while the real data indicate a much higher value of $S_\text{int} / S_\text{peak} = 1.64 \pm 0.08$. This implies that the peak flux density does not constitute an accurate measurement of the true stacked flux density. Instead, if we allow for resolved sources and random positional offsets, we find that the typical $S_\text{int}/S_\text{peak}$ we obtain for the AS2UDS sample is accurately recovered by the simulated stacks. In this case, peak flux densities underestimate the true radio flux density by a factor of $0.74\pm0.11$, which in turn leads to $q_\text{IR}$ being overestimated by a typical $\Delta q_\text{IR} = 0.13 \pm 0.06$, where the uncertainty represents the standard deviation across the stacked mock sources. However, in the SNR-range achieved for AS2UDS, we find $\Delta q_\text{IR} = -0.05 \pm 0.08$ for the integrated flux density. As such, total fluxes from Gaussian fits constitute a better measurement of the true radio flux of the AS2UDS SMGs than peak flux densities. Conversely, the measured AS2UDS stacks in the GMRT 610\,MHz map shown in Figure \ref{fig:alpha_vs_param} have an average ratio of their integrated-to-peak flux density of $S_\text{int} / S_\text{peak} = 1.07 \pm 0.05$. This value is consistent with a value of unity, and hence with the stacks being unresolved -- as may be expected given the large beam size of the GMRT observations. We therefore adopt the peak flux density for the GMRT stacks in this work. We have additionally verified through stacking of simulated sources that these peak flux densities are reliable for the GMRT data.

\begin{figure*}[!t]
    \centering
    \includegraphics[width=0.49\textwidth]{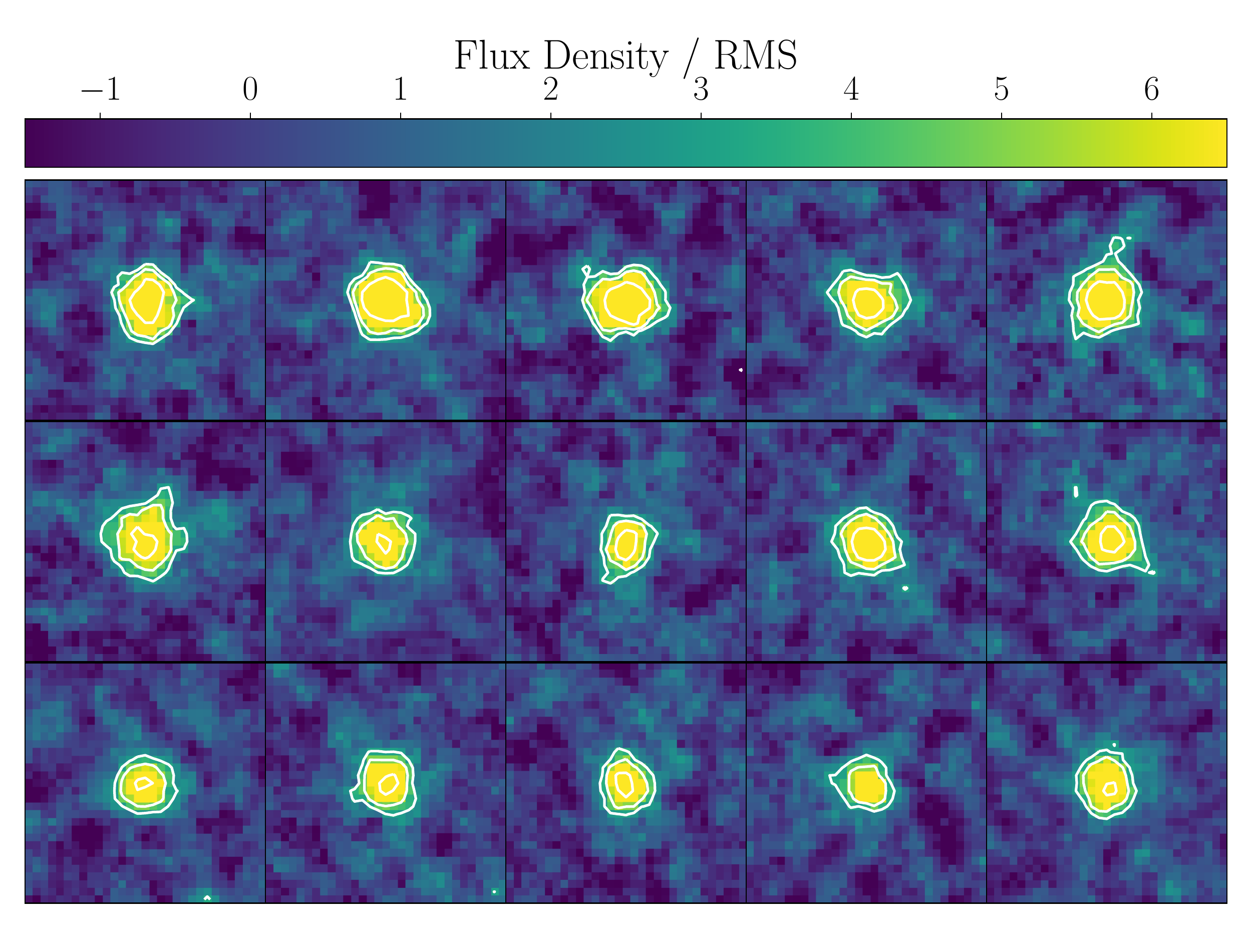}
    \includegraphics[width=0.49\textwidth]{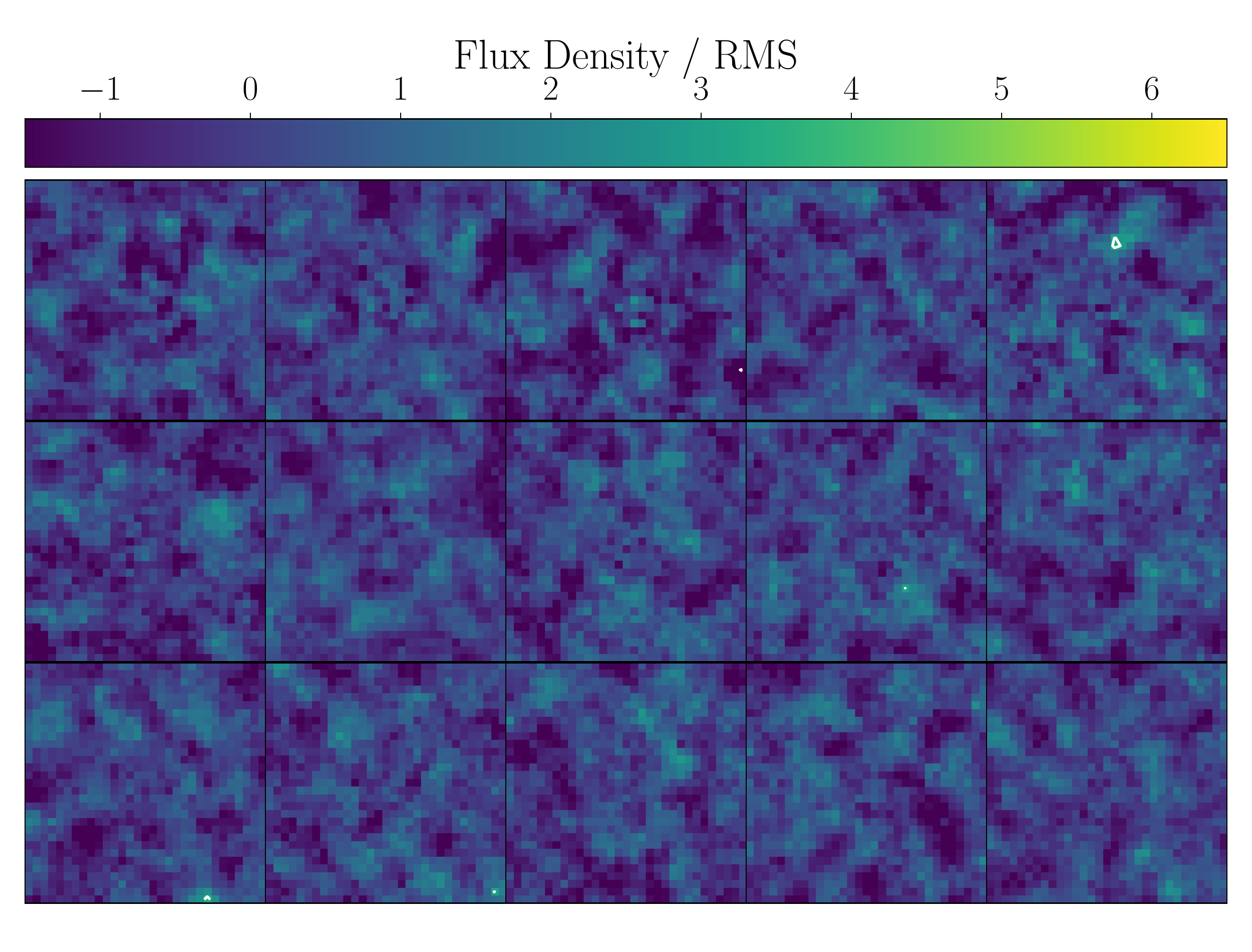} \hfill
    \caption{\textbf{Left:} Stacked detections for the 15 bins shown in Figure \ref{fig:qTIR_detections}c, combining 222 and 418 radio-detected and -undetected sources, respectively. Only the central $31\times31$ pixels ($11'' \times 11''$) are shown, for clarity. The color scheme runs from $-1.5$ to $6.5\sigma$, where $\sigma$ is the RMS-noise in the stack. Contours are shown at levels of $-3$ (dashed) and $3, 5, 9\sigma$ (solid). \textbf{Right:} The residuals, after fitting the stacked detections with an elliptical Gaussian via {\sc{PyBDSF}} \citep{mohanrafferty2015}. All stacks show a detection at a high signal-to-noise (median SNR of $11\sigma$), and are well-fit by the Gaussian model as demonstrated by the featureless residuals.}
    \label{fig:stacks}
\end{figure*}

Figure \ref{fig:mockstacking} further indicates that at low signal-to-noise, the integrated flux density becomes increasingly biased, and tends to overestimate the true flux density, similar to what was observed by \citet{leslie2020}. While we investigate this in more detail in a forthcoming paper, in this work we ensure the stacks are all of high signal-to-noise ($\text{SNR}\gtrsim10$) in order for reliable integrated flux density measurements to be made. As a representative example, we show in Figure \ref{fig:stacks} the stacks and residuals corresponding to the stacked data for the full AS2UDS sample in Figure \ref{fig:qTIR_detections}. The featureless residuals show the integrated flux density measurements to be accurate.

\section{Radio AGN in submm and radio-selected Samples}
\label{app:agn_fraction}

By construction, radio-selected studies are sensitive to the combined emission from star-formation and AGN activity, whereas sub-millimeter surveys are restricted solely to dust-obscured star formation. As such, a higher incidence of radio-excess AGN may naturally be expected in radio-selected samples.

\begin{figure*}[!t]
    \centering
    \includegraphics[width=1.0\textwidth]{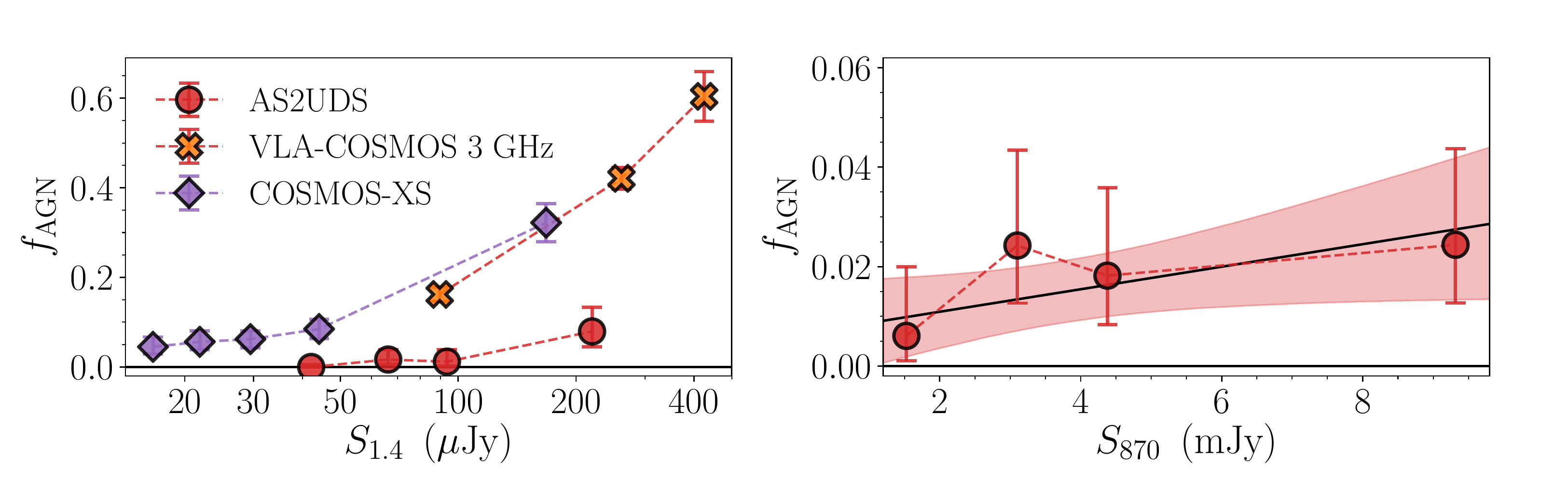}
    \caption{\textbf{Left:} The distribution of radio-excess AGN as a function of 1.4 GHz flux density. The full radio-detected AS2UDS sample is subdivided into four different bins, from which AGN are identified as having $q_\text{IR} \leq 1.55$. The uncertainty on the points represents the counting error from \citet{gehrels1986}. At flux densities $S_{1.4} \lesssim 100\,\mu$Jy, the AGN fraction is only $f_\text{AGN} = 1_{-1}^{+2} \%$. For comparison, we overplot fractions of radio-excess AGN from two deep radio surveys at 3 GHz (scaled to 1.4 GHz using $\alpha=-0.70$): COSMOS-XS (Van der Vlugt et al. 2020, Algera et al. 2020) in purple and VLA-COSMOS \citep{smolcic2017a,smolcic2017b} in orange. The incidence of radio-excess AGN in such radio-selected samples is an order of magnitude larger than in the AS2UDS SMG sample. \textbf{Right:} The distribution of radio-excess AGN as a function of $870\,\mu$m flux density, for the full AS2UDS sample, including the radio undetected population. The red, shaded region represents a linear fit through the data, and shows no evidence of a trend between between the incidence of radio AGN and sub-millimeter flux.}
    \label{fig:agn_fraction}
\end{figure*}

We show in the left panel of Figure \ref{fig:agn_fraction} the fraction of radio-excess AGN in both AS2UDS and two 3 GHz-selected radio samples (\citealt{smolcic2017b,algera2020}) as a function of radio flux density. In the latter two studies, AGN were similarly identified via a radio-excess criterion, though by adopting a redshift-dependent radio-excess threshold instead of the fixed value we adopt in this work. For the radio studies, we scale flux densities from 3 to 1.4 GHz assuming a spectral index of $\alpha = -0.70$ (following e.g., \citealt{smolcic2017a}).

At the faint end of AS2UDS, $S_{1.4} \sim 50\,\mu$Jy, \citet{algera2020} find that radio-excess AGN still make up $\sim10\%$ of the radio population, while the AGN fraction in the SMGs in this regime is consistent with zero. At higher flux densities $S_{1.4} \sim 300\,\mu$Jy, \citet{smolcic2017b} determine AGN fractions of $\sim50\%$, compared to the $\sim10\%$ for the radio-detected AS2UDS sample. This emphasizes that, while both radio and far-infrared emission are tracers of star formation, the former suffers significant contamination. The increased incidence of radio-excess AGN in radio surveys is not surprising -- after all, these constitute some of the brightest radio sources observed. However, it may suggest that studies of the FIRRC based on radio-selected samples will be biased by contamination from active galactic nuclei. In particular, such surveys will be more sensitive to composite sources, where radio emission from both AGN activity and star formation contributes to the overall radio luminosity, compared to a submm-selected sample. This is further demonstrated in the right panel of Figure \ref{fig:agn_fraction}, where we show the radio-AGN fraction as a function of $870\,\mu$m flux density. We find no evidence for any correlation, and a linear fit through the data returns a slope of $(2.2 \pm 2.3) \times 10^{-3}$, consistent with no gradient at the $1\sigma$ level. This substantiates that a FIR-based selection renders the sample insensitive to radio AGN.

\section{The FIRRC in Radio-selected Samples}
\label{app:radio_firrc}

As outlined in Section \ref{sec:comparison}, radio-selected samples typically observe a declining $q_\text{IR}$ with redshift, whereas we observe no such evolution in our sample of SMGs. A likely explanation is that this evolution towards lower values of $q_\text{IR}$ is the result of (low-level) AGN emission \citep{murphy2009b}. However, obtaining conclusive evidence for this is complicated by the fact that, by definition, these AGN are very difficult to identify in radio surveys. One clear way to distinguish AGN from purely star-forming sources in the radio, however, is through Very Large Baseline Interferometry (VLBI) observations, which are sensitive only to strong and compact sources of radio emission. If we assume the evolution in the far-infrared/radio correlation seen in recent radio surveys is solely due to low-level radio AGN, we can calculate the redshift-dependent fraction of radio emission that originates from star formation. This further requires the assumption that FIR-emission is a perfect tracer of star-formation rate across all redshifts. Given a local `correct' value of $q_\text{IR}$ equal to $q_0$, as well as redshift-evolution with slope $\gamma$, as defined in Section \ref{sec:results_qtir}, the average redshift-dependent fraction of radio emission that originates from star formation, $f_\text{SFR}$, is equal to

\begin{align}
    f_\text{SFR}(z)  = 10^{q_0 \left[(1+z)^\gamma - 1\right]} \ .
\end{align}

\citet{herreraruiz2017} observed the 2-square degree COSMOS field with VLBI observations at 1.4 GHz, reaching a sensitivity of $\sigma \sim 10 - 15\ \mu$Jy, similar to that of the original 1.4 GHz VLA COSMOS survey \citep{schinnerer2007,schinnerer2010}. They identify 438 radio sources at $\geq5.5\sigma$ significance, with a typical VLBI flux density of $\sim0.6-0.7$ times the total flux density, obtained from the lower resolution VLA observations. This constitutes a typical detection fraction of $f_\text{det} \simeq 0.20$ in the VLBI data. Additionally, they find that the sub-mJy radio population is more likely to host a dominant radio AGN, i.e. contributing a larger fraction of the total radio luminosity, though this is likely to (at least partially) be a selection effect, as faint sources require a strong AGN contribution to be detectable in the VLBI observations in the first place. In the following, we define $f_\text{AGN} = 1 - f_\text{SFR} = S_\text{VLBI} / S_\text{VLA}$, following \citet{herreraruiz2017}, and we adopt a value of $f_\text{AGN} = 0.70$, which is typical for their sample at $S_\text{VLA} \lesssim 1$mJy. Due to the aforementioned selection effects, this likely constitutes an upper limit to the actual contribution from a typical AGN within this flux density range. Additionally, we do not explicitly remove radio-excess AGN from this sample, which are typically discarded prior to calculating the FIRRC, and as a result, the calculated $f_\text{AGN}$ in this way will constitute a rather strict upper limit.

\begin{figure}[!t]
    \centering
    \hspace*{-0.3cm}\includegraphics[width=0.5\textwidth]{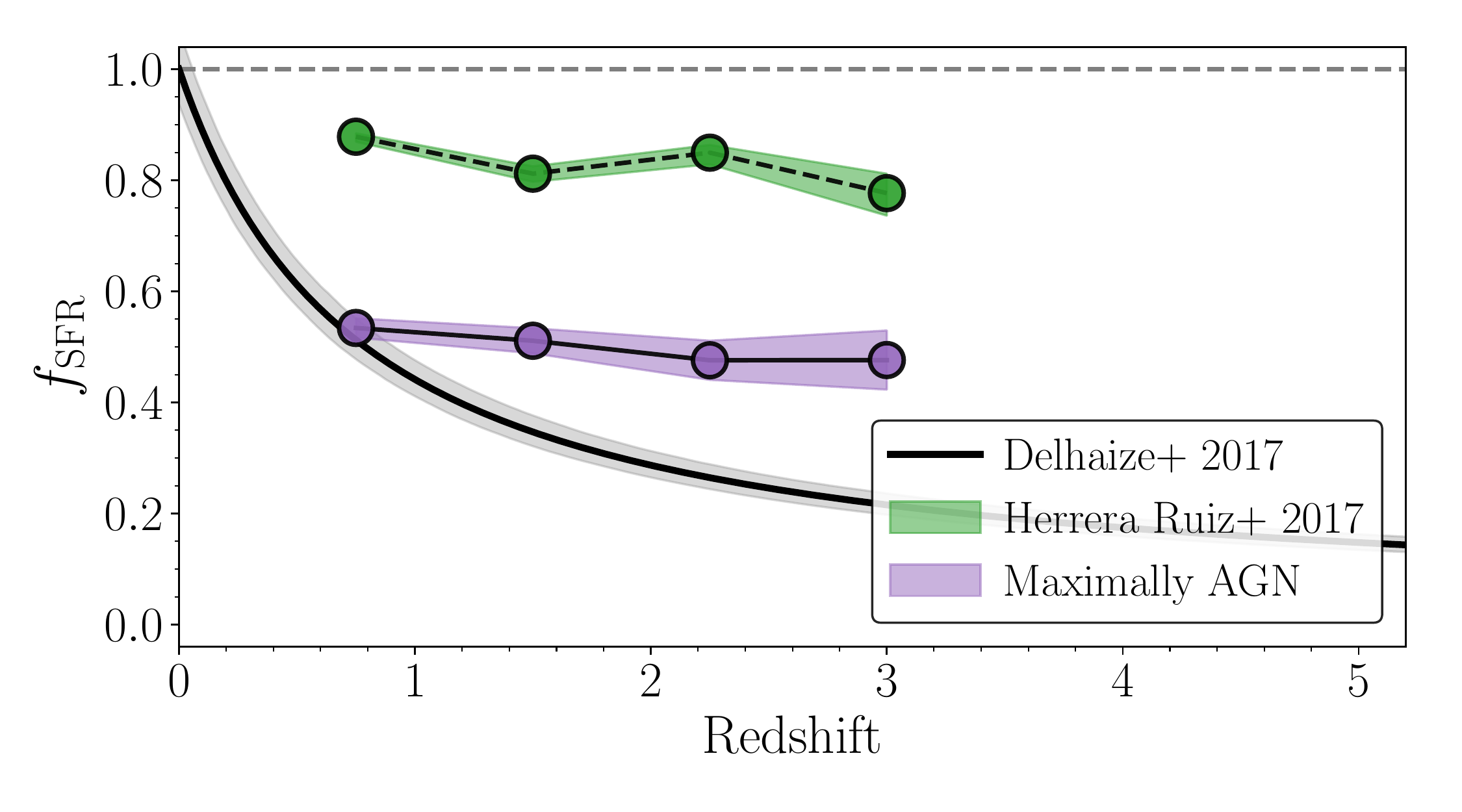}
    \caption{Fractional contribution of star-formation to the total radio emission as a function of redshift, under the assumption that the FIRRC is non-evolving, and any observed variation in the correlation is the result of emission from unidentified AGN. The black line shows the evolution measured by \citet{delhaize2017}, using data from the 3 GHz VLA-COSMOS project. The green, shaded region shows the expected fraction of star-formation-powered radio emission based on the VLBI observations from \citet{herreraruiz2017}, given their detection rate as a function of redshift, and a fixed fraction of $f_\text{AGN} = S_\text{VLBI} / S_\text{VLA} = 0.70$. This region still constitutes a lower limit on $f_\text{SFR}$ as radio-excess AGN were not removed from the VLBI sample. The purple shaded region constitutes the expected $f_\text{SFR}$ when all VLBI-undetected sources in the 1.4 GHz VLA COSMOS project \citep{schinnerer2007,schinnerer2010} have the maximum possible AGN contribution to render them just below the VLBI detection limit. Even this unrealistic scenario cannot explain the measured redshift-dependency of the FIRRC in radio studies beyond $z\gtrsim1.5$, and hence an enhanced AGN contribution at high-redshift alone is unlikely to be enough to fully account for the observed evolution in the FIRRC seen in radio-based samples.}
    \label{fig:fsfr}
\end{figure}

We show in Figure \ref{fig:fsfr} the expected $f_\text{SFR}$ as a function of redshift, assuming the evolution in the FIRRC found by \citet{delhaize2017} is solely the result of unidentified AGN contamination. We overplot the expected contribution from AGN based on the \citet{herreraruiz2017} VLBI sample, as well as the $f_\text{SFR}$ expected when \emph{all} sources from the 1.4 GHz VLA COSMOS survey that are undetected in the VLBI observations have the maximal possible AGN contribution to render them just below the detection limit, i.e.\ $f_\text{AGN}^\text{max} = 5.5\times\sigma_\text{RMS} / S_\text{VLA}$. This, evidently, constitutes a highly unrealistic scenario. The VLBI sample from \citet{herreraruiz2017} implies an upper limit on the fraction of AGN contamination on the order of $20\%$. Additionally, this fraction is not a strong function of redshift, as the fraction of known radio sources they detect in the VLBI observations does not vary much across cosmic time. While the ``maximally AGN'' scenario coincides with the expected $f_\text{SFR}$ at $z\sim1$, the lack of a redshift-dependency ensures that at higher redshifts even this worst-case scenario underestimates how much AGN are required to contribute in order to obtain the observed redshift-evolution in the FIRRC in radio-selected samples.

Overall, an enhanced AGN contribution at high-redshift alone is therefore unlikely to fully account for the observed evolution in the FIRRC seen in radio surveys, and even by requiring all VLBI-undetected sources to be radio AGN, we cannot explain the observed $f_\text{SFR}$. As a result, it is likely that the difference between the far-infrared/radio correlations for radio-selected and submm-selected samples is due to the different galaxy populations such studies probe.

\bibliographystyle{apj}
\bibliography{references}

\end{document}